\documentclass[aps,nofootinbib,showkeys,superscriptaddress,preprint]{revtex4}
\usepackage[T1]{fontenc}
\usepackage{amsmath,amssymb}
\usepackage{epsfig}
\usepackage{bbold}
\usepackage{dcolumn}
\usepackage{graphicx}
\usepackage[usenames,dvipsnames]{color}
\usepackage{slashed}
\usepackage[colorlinks,citecolor=blue]{hyperref}
\usepackage{pdfpages}
\usepackage{color}
\begin{document}
\title{Active and sterile neutrino phenomenology with $A_4$ based minimal extended seesaw}
\author{Pritam Das}
\email{pryxtm@tezu.ernet.in}
\author{Ananya Mukherjee}
\email{ananyam@tezu.ernet.in}
\author{Mrinal Kumar Das}
\email{mkdas@tezu.ernet.in}
\affiliation{%
 Department of Physics, Tezpur University, Tezpur 784\,028, India %
 }%
\begin{abstract}
We study a model of neutrino within the framework of minimal extended seesaw (MES), which plays an important role in active and sterile neutrino phenomenology in (3+1) scheme. The $A_4$ flavor symmetry is augmented by additional $Z_4\times Z_3$  symmetry to constraint the Yukawa Lagrangian of the model. We use non-trivial Dirac mass matrix, with broken $\mu-\tau$  symmetry, as the origin of leptonic mixing. Interestingly, such structure of mixing naturally leads to the non-zero reactor mixing angle $\theta_{13}$. Non-degenerate  mass structure for right-handed neutrino $M_R$ is considered so that we can further extend our study to Leptogenesis. We have also considered three different cases for sterile neutrino mass, $M_S$ to check the viability of this model, within the allowed $3\sigma$ bound in this MES framework. 
\end{abstract}

\maketitle

\section{INTRODUCTION}
Followed by the discovery of the Higgs Boson, the Standard Model (SM) of particle physics is essentially complete, although there are some insufficiencies in the theory. One needs to extend the SM in order to address phenomenon like origin of neutrino mass, dark matter, strong CP problem and matter-antimatter asymmetry, etc. Several neutrino oscillation experiments like  SK\cite{Abe:2016nxk}, SNO\cite{Boger:1999bb} MINOS\cite{Evans:2013pka}, T2K\cite{Abe:2011sj}, RENO\cite{Ahn:2012nd}, DOUBLE CHOOZ\cite{Abe:2011fz}, DAYABAY\cite{An:2012eh}, etc. have established the fact that neutrinos produced in a well-defined flavor eigenstate can be detected as a different flavor eigenstate while they propagate. This can be interpreted as, like all charged fermions, neutrinos have mass and mixing because their flavor eigenstates are different from mass eigenstates. The existence of neutrino mass was the first evidence for the new physics beyond the Standard Model (BSM). Some recent reviews on neutrino physics are put into references \cite{Mohapatra:2005wg,King:2003jb,King:2014nza}.

In standard neutrino scenario three active neutrinos are involved with two mass square differences\footnote{order of $ 10^{-5}eV^{2} $ and  $ 10^{-3}eV^{2} $ for solar ($\Delta m^{2}_{21}$) and atmospheric ($\Delta m^{2}_{23}/\Delta m^{2}_{13}$) neutrino respectively.}, three mixing angles ($\theta_{ij};i,j=1,2,3$) and one Dirac CP phase ($\delta_{13}$). Earlier it was assumed that the reactor mixing angle $\theta_{13}$ is zero but later in 2012 it was measured with incredible accuracy: $\theta_{13} \sim 8.5^0 \pm0.2^0$ \cite{An:2012eh}. If neutrinos are Majorana particles then there are two more CP violating phases ($\alpha$ and $\beta$) come into the 3-flavor scenario. Majorana phases are not measured experimentally as they do not involve in the neutrino oscillation probability. The current status of global analysis of neutrino oscillation data \cite{Gonzalez-Garcia:2015qrr,Forero:2014bxa,Capozzi:2016rtj} give us the allowed values for these parameters in $3\sigma$ confidence level, which is shown in Table \ref{tab:d1}. Along with the Majorana phases, the absolute mass scale for the individual neutrino is still unknown as the oscillation experiments are only sensitive to the mass square differences, even though Planck data constrained the sum of the three neutrinos, $\Sigma m_{\nu}<0.17eV$ at $95{\%}$ confidence level \cite{Palanque-Delabrouille:2015pga}. Due to the fact that absolute scale of the neutrino mass is not known yet, as the oscillation probability depends on the mass square splittings but not the absolute neutrino mass. Moreover, neutrino oscillation experiments tell that the solar mass square splitting is always positive, which implies $m_2$ is always grater than $m_1$. However, the same confirmation we have not yet received regarding the atmospheric mass square splittings from the experiments. This fact allows us to have two possible mass hierarchy patterns for neutrinos; Normal Hierarchy (NH:$m_1\ll m_2<m3$) as well as Inverted Hierarchy (IH:$m_3\ll m_1<m_2$). 

In past few decades, there has been successful achievements in solar, reactor and accelerator experiments whose results are in perfect agreement with only three active neutrino scenario meanwhile there are some anomalies which need explanation. The very first and most distinguished results towards new physics in the neutrino sector were from LSND results \cite{Athanassopoulos:1997pv,Aguilar:2001ty,Athanassopoulos:1996jb}, where electron anti-neutrino ($\overline{\nu_{e}}$) were observed in the from of muon anti-neutrino ($\overline{\nu_{\mu}}$) beam seemingly $\overline{\nu_{e}}$ was originally $\overline{\nu_{\mu}}$. Moreover, data from MiniBooNE \cite{Aguilar-Arevalo:2012fmn} results overlap with LSND results and give an indication towards extra neutrino hypothesis. To make sure that these data are compatible with current picture one needs new mass eigenstates for neutrinos. These additional states must relate to right-handed neutrinos (RHN) for which bare mass term are allowed by all symmetries i.e. they should not be present in $SU(2)_{L}\times U(1)_{Y}$ interactions, hence are \textit{Sterile}. Recently observed Gallium Anomaly observation\cite{Abdurashitov:2005tb, Giunti:2010zu, Giunti:2012tn} is also well explained by sterile neutrino hypothesis. Although there are few talks about the non-existence of extra neutrino, but finally reactor anti-neutrino anomaly results \cite{Mention:2011rk, Kopp:2011qd} give a clear experimental proof that the presence of this fourth non-standard neutrino is mandatory. Moreover, cosmological observation \cite{Hamann:2010bk} (mainly CMB\footnote{cosmic microwave background} or SDSS\footnote{Sloan Digital Sky Survey}) also favor the existence of sterile neutrino. From cosmological consequences, it is said that the sterile neutrino has a potential effect on the entire Big-Bang Nucleosynthesis \cite{Izotov:2010ca}. LSND results predicted sterile neutrino with mass $\sim\mathcal{O}(1)eV$. To be more specific with the recent update with MiniBooNE experiment results \cite{Aguilar-Arevalo:2018gpe}, which combine the $\nu_e/\overline{\nu_{e}}$ appearance data with the LSND results to establish the presence of an extra flavor of neutrino upto 6.0$\sigma$ confidence level. However these results from LSND/MiniBooNE are in tension with improved bounds on appearance/disappearance experiments results from IceCube/MINOS+ \cite{Dentler:2018sju}. Further discussion on this argument is beyond the scope of this paper. Although $\Delta{m_{41}^{2}}\sim1eV^2$ is consistent with global data from the $\nu_{e}$ disappearance channel which supports sterile neutrino oscillation at $3\sigma$ confidence level. Thus, hints from different backgrounds point a finger towards the presence of a new generation of neutrinos. 
 
Sterile neutrino is a neutral lepton which does not involve itself in weak interactions, but they are induced by mixing with the active neutrinos that can lead to observable effect in the oscillation experiments. Furthermore, they could interact with gauge bosons which lead to some significant correction in non-oscillation processes e.g., in the neutrinoless double beta decay (NDBD) amplitude\cite{Goswami:2005ng,Goswami:2007kv}, beta decay spectra. Since RH neutrinos are SM gauge singlets\cite{deGouvea:2005er}, so it is possible that sterile neutrinos could fit in the canonical type-I seesaw as the RH neutrino if their masses lie in the eV regime. Some global fit studies have been carried out for sterile neutrinos at eV scale being mixed with the active neutrinos \cite{Kopp:2013vaa,Giunti:2013aea,Gariazzo:2015rra}. While doing this the Yukawa Coupling relating lepton doublets and right-handed neutrinos should be of the order $10^{-12}$ which implies a Dirac neutrino mass of sub-eV scale to observe the desired active-sterile mixing. These small Dirac Yukawa couplings are considered unnatural unless there is some underlying mechanism to follow. Thus, it would be captivating to choose a framework which gives low-scale sterile neutrino masses without the need of Yukawa coupling and simultaneously explain active-sterile mixing. In order to accommodate sterile neutrino in current SM mass pattern, various schemes were studied. In (2+2) scheme, two different classes of neutrino mass states differ by $eV^2$, which is disfavored by current solar and atmospheric data \cite{Maltoni:2002ni}. Current status for mass square differences, corresponding to sterile neutrinos, dictates sterile neutrinos to be either heavier or lighter than the active ones. Thus, we are left with either (1+3) or (3+1) scheme. In the first case, three active neutrinos are in eV scale and sterile neutrino is lighter than the active neutrinos. However, this scenario is ruled out by cosmology \cite{Hamann:2010bk,Giusarma:2011ex}. In the latter case, three active neutrinos are in sub-eV scale and sterile neutrino is in eV scale \cite{GomezCadenas:1995sj,Goswami:1995yq}. Numerous studies have been exercised taking this (3+1) framework with various prospects \cite{Giunti:2011gz,Borah:2016xkc,Kopp:2011qd,Gariazzo:2015rra}.

 The seesaw mechanism is among one of the most prominent theoretical mechanism to generate light neutrino masses naturally. Various types of see-saw mechanisms have been put in literature till date (for detail one may look at \cite{minkowski77,mohapatra79,Weinberg:1979sa,Babu:1993qv,Mukherjee:2015axj,Dev:2012bd,King:2003jb,schechter81,foot88,ma98,bernabeu87,tHooft:1980xss,mohapatra86}). In our study, we will focus our model to fit with (3+1) framework where the sterile neutrino is in the eV range and the active neutrinos in sub-eV range. Study of eV sterile neutrino in Flavor symmetry model have been discussed by various authors in \cite{Altarelli:2009kr,Barry:2011wb,Chun:1995bb,Chen:2011ai,deGouvea:2006gz}. There has been plenty of exercises performed in order to study eV scale sterile neutrino phenomenology through the realization of Froggatt-Nielsen (FN) mechanism\cite{Froggatt:1978nt} adopting non-Abelian $A_4$ flavor symmetry in seesaw framework \cite{Barry:2011wb,Zhang:2011vh,Heeck:2012bz,Babu:2009fd}. Similar approaches using type-I seesaw framework have been evinced  by some authors\cite{Zhang:2011vh,Barry:2011wb,Nath:2016mts}, where type-I seesaw is extended by adding one extra singlet fermion, which scenario is popularly known as the minimal extended seesaw (MES) model. This extension gives rise to tiny active neutrino mass along with the sterile mass without the need of small Yukawa couplings. There are few literature available for model termed as $\nu MSM$ \cite{Asaka:2005pn, Shaposhnikov:2008pf}, where SM is extended using three right handed (RH) neutrino (with masses smaller than electroweak (EW) scale), which is a simplest and most economical extension of Standard Model to explain $keV$ scale sterile neutrino with other BSM phenomenons. Our considered MES framework is more or less analogous to the $\nu MSM$ framework which is also extended with three RH neutrinos along with a chiral singlet. However within $\nu MSM$ sterile neutrino mass scale is fixed within $keV$ range while in with MES we can tune the range of the sterile neutrino mass from $eV$ scale to $keV$ scale. This unique feature of MES encourages us to study sterile neutrino over the $\nu MSM$ framework. Parallel with the MES and $\nu MSM$, inverse seesaw (ISS) framework is quite popular in literature to study sterile neutrino phenomenology\cite{Dev:2012bd,Boulebnane:2017fxw}. Although in this work we are quite focused with MES framework to study sterile neutrino, however for readers choice, a generalized comparison between ISS and Extended seesaw is provided in the appendix \ref{apa}.  
 
 In this paper, we have studied the active and sterile neutrino mixing scheme within the MES framework based on $A_4$ flavor symmetry along with the discrete $Z_4$ and $Z_3$ symmetry.  There are few works on MES based on $A_4$  are available in literature \cite{Barry:2011wb,Zhang:2011vh}. Those studies were carried out prior to the discovery of non-zero reactor mixing angle $\theta_{13}$. In our model, we have considered different flavons to construct the non-trivial Dirac mass matrix ($M_D$), which is responsible for generating light neutrino mass. In this context, we have added a leading order correction to the Dirac mass matrix to accumulate non-zero reactor mixing angle($\theta_{13}$), in lieu of considering higher order correcting term in the Lagrangian as mentioned in the ref. \cite{Altarelli:2005yp}.
 As mentioned, in spite of having $M_D,\ M_R,\ M_S$ matrices, we have introduced the new leading order correction matrix $M_P$ which is produced from a similar kind of coupling term that accomplish the Dirac mass matrix ($M_D$). $M_P$ is added to $M_D$, such that there is a broken $\mu-\tau$ symmetry which leads to the generation of the non-zero reactor mixing angle. The $M_D$ matrix constructed for NH does not work for IH, the explanation of which we have given in the model section. Thus, we have reconstructed $M_D$ by introducing a new flavon ($\varphi^{\prime}$) to the Lagrangian to study the case of IH pattern. A most general case also has been introduced separately where the non-zero $\theta_{13}$ is automatically generated by a different $M_D$ constructed with the help of a most general kind of VEV alignment. 
A non-degenerate mass structure is considered for the diagonal $M_R$ matrix so that we can extend our future study towards Leptogenesis.  As mentioned, we find two such frameworks very appealing where neutrino masses considered to be of $\mu-\tau$ symmetric type \cite{Barry:2011wb,Zhang:2011vh}. But here in our work we have extensively studied the consequences brought out by taking  the sterile mass pattern via altering the position for the non-zero entry in $M_S$. All these $M_S$ structures have been studied independently for both the mass ordering and results are plotted in section \ref{sec4}. In the phenomenology part, we have constrained the model parameters in the light of current experimental data and also shown correlation between active and sterile mixing by considering three different $M_S$ structures.

\begin{table}
\begin{center}
\begin{tabular}{||c|c|c||}
\hline
Parameters&NH (Best fit)& IH (Best fit)\\
\hline
$\Delta m^{2}_{21}[10^{-5}eV^2]$&6.93-7.97(7.73)&6.93-7.97(7.73)\\ \hline
$\Delta m^{2}_{31}[10^{-3}eV^2]$&2.37-2.63(2.50)&2.33-2.60(2.46)\\ \hline
$sin^{2}\theta_{12}/10^{-1}$&2.50-3.54(2.97)&2.50-3.54(2.97)\\ \hline
$sin^{2}\theta_{13}/10^{-2}$&1.85-2.46(2.14)&1.86-2.48(2.18)\\ \hline
$sin^{2}\theta_{23}/10^{-1}$&3.79-6.16(4.37)&3.83-6.37(5.69)\\ \hline
$\delta_{13}/\pi$&0-2(1.35)&0-2(1.32)\\ \hline
$\Delta m^{2}_{LSND}(\Delta m^{2}_{41} \text{or} \Delta m^{2}_{43})eV^2$&0.87-2.04(1.63)&0.87-2.04(1.63)\\ \hline
$|V_{e4}|^{2}$&0.012-0.047(0.027)&0.012-0.047(0.027)\\ \hline
$|V_{\mu4}|^{2}$&0.005-0.03(0.013)&0.005-0.03(0.013)\\ \hline
$|V_{\tau4}|^{2}$&<0.16(-)&<0.16(-)\\ \hline

\end{tabular}

\caption{The latest global fit $3\sigma$ range and best fit results from recent active neutrino parameters\cite{Capozzi:2016rtj}. The current sterile neutrino bounds are from  \cite{Gariazzo:2015rra,Nath:2016mts}.}\label{tab:d1}
\end{center}
\end{table}
This paper is organized as follows. In section \ref{sec2} brief review of the minimal extended seesaw is given. In section \ref{sec3} we have discussed the $A_4$ model and generation of the mass matrices in the leptonic sector. We keep the section \ref{sec4} and its subsections for numerical analysis in NH and IH case respectively. Finally, the summary of our work is concluded in the section \ref{sec5}.

\section{THE MINIMAL EXTENDED SEESAW}
\label{sec2}
In the present work we have used Minimal extended seesaw(MES) which enable us to connect active neutrino with sterile neutrino of a wider range\cite{Barry:2011wb}. In this section, we describe the basic structure of MES, where canonical type-I seesaw is extended to achieve eV-scale sterile neutrino without the need of putting tiny Yukawa coupling or any small mass term. In MES scenario along with the SM particle, three extra right-handed neutrinos and one additional gauge singlet chiral field S is introduced. The Lagrangian of the neutrino mass terms for MES is given by: 
\begin{equation}
-\mathcal{L}_{\mathcal{M}}= \overline{\nu_{L}}M_{D}\nu_{R}+\frac{1}{2}\overline{\nu^{c}_{R}}M_{R}\nu_{R}+\overline{S^c}M_{S}\nu_{R}+h.c. ,
 \end{equation} 
  The neutrino mass matrix will be a $7\times7$ matrix, in the basis ($\nu_{L},\nu_{R}^{c},S^c$), reads as
 \begin{equation}
 M_{\nu}^{7\times7}=
 \begin{pmatrix}
0&M_D&0\\M_{D}^{T}&M_{R}&M_{S}^{T}\\0&M_{S}&0 
 \end{pmatrix}\label{7b7}.
 \end{equation}
   Here $M_D$ and $M_R$ are $3\times3$ Dirac and Majorana mass matrices respectively whereas $M_S$ is a $1\times3$ matrix. As per the standard argument \cite{Schechter:1980gr} the number of massless state is defined as $n(\nu_L)+n(S)-n(\nu_R)$, within our framework it is one ($3+1-3=1$), which is also verified by taking the determinant of the matrix \eqref{4b4} in the next paragraph of this section. The zeros at the corners of the $7\times7$ matrix of \eqref{7b7} has been enforced and motivated by some symmetry. This can be achieved with discrete flavor symmetry due to which it is clear that right handed neutrinos and $S$ carry different charges. Moreover, the MES structure could also be explained with the abelian symmetry. For example, one may introduce additional $U(1)^{\prime}$ under which along with the all SM particles we assumed and 3 RH neutrinos to be neutral. The RH singlet $S$ on the other hand carries a $U(1)^{\prime}$ charge $Y^{\prime}$ and we further introduce a SM singlet $\chi$ with hypercharge $-Y^{\prime}$. The matrix $M_S$ is generated by the gauge invariant coupling $S^c\chi\nu_R$ after $\chi$ acquires a VEV, while the Majorana mass for $S$($i.e., \overline{S^c}S$) and a coupling with the active neutrino $\nu_L$ are still forbidden by the $U(1)^{\prime}$ symmetry at the renormalizable level\cite{Barry:2011wb,Chun:1995js, Heeck:2012bz}. This explains the zeros in the $7\times7$ matrix. 
 In the analogy of type-I seesaw the mass spectrum of these mass matrices are considered as $M_{R}\gg M_{S}>M_{D}$, so that the heavy neutrinos decoupled at low scale. After diagonalizing, $4\times4$ neutrino mass matrix in the basis $(\nu_{L},S^c)$, is given by,
 \begin{equation}
 M_{\nu}^{4\times4}=
 -\begin{pmatrix}
 M_{D}M_{R}^{-1}M_{D}^{T}&M_{D}M_{R}^{-1}M_{S}^{T}\\
 M_{S}(M_{R}^{-1})^{T}M_{D}^{T} & M_{S}M_{R}^{-1}M_{S}^{T}
 \end{pmatrix}\label{4b4}.
 \end{equation}
 Here in  $M_{\nu}^{4\times4}$ matrix \eqref{4b4}, there are three eigenstates exists for three active neutrinos and one for the light sterile neutrino. Taking the determinant of Eq.(3), we get,
 \begin{equation}
 \begin{split}
  \text{det}(M^{4\times4}_{\nu})& =
 \text{det}(M_{D}M_{R}^{-1}M_{D}^{T})\text{det}[ -M_{S}M_{R}^{-1}M_{S}^{T}+ M_{S}(M_{R}^{-1})^{T}M_{D}^{T}( M_{D}M_{R}^{-1}M_{D}^{T})^{-1}(M_{D}M_{R}^{-1}M_{S}^{T})]\\
& =\text{det}(M_{D}M_{R}^{-1}M_{D}^{T})\text{det}[M_{S}(M_{R}^{-1}-M_{R}^{-1})M_{S}^{T}]\\
& =0.
 \end{split}
 \end{equation}
 Here the zero determinant indicates that one of the eigenvalue is zero. Thus, the MES formalism demands one of the light neutrino mass be exactly vanished.\\
 Proceeding for diagonalization, we face three choices of ordering of $M_S$ :
\begin{itemize}
	\item $M_{D}\sim M_{S}$: This indicates a maximal mixing between active and sterile neutrinos which is not compatible with the neutrino data.
	\item $M_{D}>M_{S}$: The light neutrino mass is obtained same as type-I seesaw i.e., $m_{\nu}\simeq-M_{D}M_{R}^{-1}M_{D}^{T}$ and the sterile neutrino mass is vanishing. Moreover, from the experimental active-sterile mass squared difference result, the active neutrino masses would be in the eV scale which would contradict the standard Planck limit for the sum of the active neutrinos.\\
	 Finally, we have the third choice,
	 \item $M_{S}>M_{D}$: which would give the possible phenomenon for active-sterile mixing.
\end{itemize}
 Now applying the seesaw mechanism to Eq. \eqref{4b4}, we get the active neutrino mass matrix as
 \begin{equation}\label{amass}
 m_{\nu}\simeq M_{D}M_{R}^{-1}M_{S}^T(M_{S}M_{R}^{-1}M_{S}^{T})^{-1}M_{S}(M_{R}^{-1})^{T}M_{D}^{T}-M_{D}M_{R}^{-1}M_{D}^{T}, 
 \end{equation} and the sterile neutrino mass as
 \begin{equation}\label{smass}
 m_{s}\simeq -M_{S}M_{R}^{-1}M_{S}^{T}.
 \end{equation}
 The first term of the active neutrino mass does not vanish since $M_S$ is a vector rather than a square matrix. It would lead to an exact cancellation between the two terms of the active neutrino mass term if $M_S$ were a square matrix. In our study the $m_s$ mass scale is in eV range whereas $M_S$ scale is slightly greater than $M_D$ scale, which is near to EW scale.
\section{THE MODEL}\label{sec3}
\subsection{Normal Hierarchy}
Non-Abelian discrete flavor symmetry like $A_4, S_4$ etc. along with $Z_n$ have played an important role in particle physics. In particular, $A_4$ is more popular in literature in explaining neutrino mass \cite{Altarelli:2009kr,Barry:2011wb,Chun:1995bb,Chen:2011ai,deGouvea:2006gz,Zhang:2011vh,Heeck:2012bz,Felipe:2013vwa,Babu:2009fd,Ma:2009wi}. $A_4$ being the discrete symmetry group of rotation leaving a tetrahedron invariant. It has 12 elements and 4 irreducible representation denoted by $\bf{1},\bf{1^{\prime}},\bf{1^{\prime\prime}} $ and $\bf{3}$. The product rules for these representations are given in appendix \ref{a4p}. Our present work is an extension of $A_4\times Z_4\times Z_3$ flavor symmetry. Here, we have assigned left-handed (LH) lepton doublet $l$ to transform as $A_4$ triplet whereas right-handed (RH) charged leptons ($e^c,\mu^c,\tau^c$) transform as 1,$1^{\prime\prime}$ and $1^{\prime}$ respectively. The flavor symmetry is broken by the triplets $\zeta, \varphi $ and two singlets $\xi$ and $\xi^{\prime}$. Besides the SM Higgs $H$, we have introduced two more Higgs ($H^{\prime},H^{\prime\prime}$)\cite{Felipe:2013vwa,Nath:2016mts} which remain invariant under $A_4$. We also have restricted non-desirable interactions while constructing the mass matrices. The particle content and the  $A_4 \times Z_4 \times Z_3$ charge assignment are shown in the table \ref{tab1}.


\begin{table}
\begin{tabular}{||c|cccc|ccc|cccc|ccc||}
\hline
Field & $l$ &$e_{R}$&$\mu_{R}$&$\tau_{R}$&$H$&$H^{\prime}$&$H^{\prime\prime}$&$\zeta$&$\varphi$&$\xi$&$\xi^{\prime}$&$\nu_{R1}$&$\nu_{R2}$&$\nu_{R3}$\\
\hline

SU(2)&2&1&1&1&2&2&2&1&1&1&1&1&1&1\\
\hline
$A_4$&3&1&$1^{\prime\prime}$&$1^{\prime}$&1&1&1&3&3&1&$1^{\prime}$&1&$1^{\prime}$&1\\
\hline
$Z_4$&1&-1&-1&-1&1&i&-1&-1&1&1&-1&1&-i&-1\\
\hline
$Z_3$&1&1&1&1&1&$\omega$&1&1&1&1&$\omega^2$&1&$\omega^2$&1\\
\hline
\end{tabular}
\caption{Particle content and their charge assignments under SU(2),$A_4$ and $Z_4\times Z_3$ groups.}\label{tab1} 
\end{table}

\begin{table}
\begin{tabular}{||c|c|c|c|c||}
\hline
Charges & $S_1$&$S_2$&$S_3$&$\chi$\\
\hline
$A_4$&$1^{\prime\prime}$&$1^{\prime}$&$1^{\prime\prime}$&$1^{\prime}$\\
\hline
$Z_4$&-i&1&i&i\\
\hline
$Z_3$&1&$\omega$&1&1\\
\hline
\end{tabular}
\caption{Scalar singlet fields and their transformation properties under $A_4$ and $Z_4\times Z_3$ groups.}\label{tab2}
\end{table}

 The leading order invariant Yukawa Lagrangian for the lepton sector is given by,
\begin{equation}
\mathcal{L} = \mathcal{L}_{\mathcal{M_\iota}}+\mathcal{L}_{\mathcal{M_D}}+\mathcal{L}_{\mathcal{M_R}}+\mathcal{L}_{\mathcal{M}_S}+h.c. .
\end{equation}
Where,
\begin{equation}
\begin{split}
\mathcal{L}_{\mathcal{M_\iota}} &= \frac{y_{e}}{\Lambda}(\overline{l} H \zeta)_{1}e_{R}+\frac{y_{\mu}}{\Lambda}(\overline{l} H \zeta)_{1^{\prime}}\mu_{R}+\frac{y_{\tau}}{\Lambda}(\overline{l} H \zeta)_{1^{\prime\prime}}\tau_{R}, \\
\mathcal{L}_{\mathcal{M_D}}= & \frac{y_{1}}{\Lambda}(\overline{l}\tilde{H}\varphi)_{1}\nu_{R1}+\frac{y_{2}}{\Lambda}(\overline{l}\tilde{H^{\prime}}\varphi)_{1^{\prime\prime}}\nu_{R2}+\frac{y_{3}}{\Lambda}(\overline{l}\tilde{H^{\prime\prime}}\varphi)_{1}\nu_{R3},\\
\mathcal{L}_{\mathcal{M_R}}= & \frac{1}{2}\lambda_{1}\xi\overline{\nu^{c}_{R1}}\nu_{R1}+\frac{1}{2}\lambda_{2}\xi^{\prime}\overline{\nu^{c}_{R2}}\nu_{R2}+\frac{1}{2}\lambda_{3}\xi\overline{\nu^{c}_{R3}}\nu_{R3}.\\
\end{split}
\end{equation}
We have extended our study with three variety of $M_S$ structures, which is generated by the interaction of a singlet field $S_i$ and the right-handed neutrino $\nu_{Ri}$. The $A_4\times Z_4$ charge alignment for the scalar fields are given in table \ref{tab2}. The effective mass term for each of the above three cases are as follows, 
\begin{equation}
\begin{split}
\mathcal{L}_{\mathcal{M}_{S}^{1}}=& \frac{1}{2}\rho\chi\overline{S_1^{c}}\nu_{R1} ,\\
\mathcal{L}_{\mathcal{M}_{S}^{2}}=& \frac{1}{2}\rho\chi\overline{S_2^{c}}\nu_{R2} ,\\
\mathcal{L}_{\mathcal{M}_{S}^{3}}=& \frac{1}{2}\rho\chi\overline{S_3^{c}}\nu_{R3} .\\
\end{split}
\end{equation}

 In the Lagrangian, $\Lambda$ represents the cut-off scale of the theory, $y_{\alpha,i}$, $\lambda_{i}$ (for $\alpha=e,\mu,\tau$ and $i=1,2,3$) and $\rho$ representing the Yukawa couplings for respective interactions and all Higgs doublets are transformed as $\tilde{H} = i\tau_{2}H^*$ (with $\tau_{2}$ being the second Pauli's spin matrix)  to keep the Lagrangian gauge invariant. Following VEV alignments of the extra flavons are required to generate the desired light neutrino mas matrix\footnote{ A discussion on minimization of VEV alignment for the triplet fields($\zeta$ and $\varphi$) is added in the appendix section.}.
\begin{equation*}
\begin{split}
&\langle \zeta \rangle=(v,0,0),\\
& \langle\varphi\rangle=(v,v,v),\\
& \langle\xi\rangle=\langle\xi^{\prime}\rangle=v,\\
& \langle\chi\rangle=u.
\end{split}
\end{equation*}
Following the $A_4$ product rules and using the above mentioned VEV alignment, one can obtain the charged lepton mass matrix as follows,
\begin{equation}
M_{l} = \frac{\langle H\rangle v}{\Lambda}\text{diag}(y_{e},y_{\mu},y_{\tau}).
\end{equation}
The Dirac\footnote{ $M_D^{\prime}$ represents the uncorrected Dirac mass matrix which is unable to generate $\theta_{13}\neq 0$. The corrected $M_D$ is given by equation \eqref{md} } and Majorana neutrino mass matrices are given by,
\begin{equation}
M^{\prime}_{D}=
\begin{pmatrix}
a&b&c\\
a&b&c\\
a&b&c\\
\end{pmatrix},M_{R}=\begin{pmatrix}
d&0&0\\
0&e&0\\
0&0&f\\
\end{pmatrix};
\label{emd}
\end{equation}
where, $a=\frac{\langle H\rangle v}{\Lambda}y_{1} , b=\frac{\langle H\rangle v}{\Lambda}y_{2} $ and $c=\frac{\langle H\rangle v}{\Lambda}y_{3}$. The elements of the $M_R$ are defined as $d=\lambda_{1}v, e=\lambda_{2}v$ and $f=\lambda_{3}v$. \\
Three different structures for $M_{S}$ reads as,   
\begin{equation}
M_{S}^{1}= \begin{pmatrix}
g&0&0\\
\end{pmatrix},\;
M_{S}^{2}= \begin{pmatrix}
0&g&0\\
\end{pmatrix}\
\text{and} \
M_{S}^{3}= \begin{pmatrix}
0&0&g\\
\end{pmatrix}.
\end{equation}
Considering only $M_{S}^{1}$ structure, the light neutrino mass matrix takes a symmetric form as,
\begin{equation}\label{symm}
m_{\nu}= \begin{pmatrix}
-\frac {b^2} {e}-\frac{c^2}{f} &-\frac {b^2} {e}-\frac{c^2}{f} &-\frac {b^2} {e}-\frac{c^2}{f} \\
-\frac {b^2} {e}-\frac{c^2}{f} &-\frac {b^2} {e}-\frac{c^2}{f}&-\frac {b^2} {e}-\frac{c^2}{f} \\
-\frac {b^2} {e}-\frac{c^2}{f}&-\frac {b^2} {e}-\frac{c^2}{f}&-\frac {b^2} {e}-\frac{c^2}{f}\\
\end{pmatrix}.
\end{equation}
As we can see, this $m_{\nu}$\footnote{ We have used $M_D^{\prime}$ in lieu of $M_D$ and $M_S^1$ instead of $M_S$ in equation \eqref{amass}} is a symmetric matrix (Democratic) generated by $M_D^{\prime},\ M_R$ and $M_S^1$ matrices. It can produce only one mixing angle and one mass square difference. This symmetry must be broken in order to generate two mass square differences and three mixing angles. In order to introduce $\mu-\tau$ asymmetry in the light neutrino mass matrix we introduce a new $SU(2)$ singlet flavon field ($\eta$), the coupling of which give rise to a matrix \eqref{pmatrix} which later on makes the matrix \eqref{symm} $\mu-\tau$ asymmetric after adding \eqref{pmatrix} to the Dirac mass matrix($M_D^{\prime}$). This additional flavon and thereby the new matrix \eqref{pmatrix} have a crucial role to play in reproducing nonzero reactor mixing angle. The Lagrangian responsible for generating the matrix \eqref{pmatrix} can be written as, 
\begin{equation}
\mathcal{L}_{\mathcal{M_P}} = \frac{y_{1}}{\Lambda}(\overline{l}\tilde{H} \eta)_1 \nu_{R1}+\frac{y_{2}}{\Lambda}(\overline{l} \tilde{H^{\prime}}\eta)_{1^{\prime\prime}}\nu_{R2}+\frac{y_{3}}{\Lambda}(\overline{l}\tilde{H^{\prime\prime}} \eta)_{1^{\prime}}\nu_{R3}.
\end{equation}
The singlet flavon field ($\eta$) is supposed to take $A_4\times Z_4 \times Z_3$ charges as same as $\varphi$ (as shown in the table \ref{tab1}). Now, considering VEV for the new flavon field as  $\langle \eta \rangle=(0,v,0)$, we get the matrix as, 
\begin{equation}\label{pmatrix}
M_{P}=
\begin{pmatrix}
0&0&p\\
0&p&0\\
p&0&0\\
\end{pmatrix}.
\end{equation}
Hence $M_D$ from eq. \eqref{emd} will take new structure as,
\begin{equation} \label{md}
	M_D=M^{\prime}_D+M_P=
	\begin{pmatrix}
	a&b&c+p\\
	a&b+p&c\\
	a+p&b&c\\
	\end{pmatrix}.
	\end{equation}
\subsection{Inverted Hierarchy} 
 Earlier in the work\cite{Zhang:2011vh}, author have explained the necessity of a new flavon in order to realize the IH within the MES framework. In our present work, we also have modified the Lagrangian for the $M_D$ matrix by introducing a new triplet flavon $\varphi^{\prime}$ with VEV alignment as $\langle\varphi^{\prime}\rangle \sim (2v,-v,-v)$, which affects only the Dirac neutrino mass matrix and give desirable active-sterile mixing in IH. The invariant Yukawa Lagrangian for the $M_D$ matrix will be,
\begin{equation}
\mathcal{L}_{\mathcal{M_D}}= \frac{y_{1}}{\Lambda}(\overline{l}\tilde{H_1}\varphi)_{1}\nu_{R1}+\frac{y_{2}}{\Lambda}(\overline{l}\tilde{H_2}\varphi^{\prime})_{1^{\prime\prime}}\nu_{R2}+\frac{y_{3}}{\Lambda}(\overline{l}\tilde{H_3}\varphi)_{1}\nu_{R3}.\\
\end{equation}
Hence the Dirac mass matrix will have the form, 
\begin{equation}
M^{\prime}_{D}=
\begin{pmatrix}
a&-b&c\\
a&-b&c\\
a&2b&c\\
\end{pmatrix},
\end{equation}
with, $a=\frac{\langle H\rangle v}{\Lambda}y_{1} , b=\frac{\langle H\rangle v}{\Lambda}y_{2} $ and $c=\frac{\langle H\rangle v}{\Lambda}y_{3}$. 

This Dirac mass matrix will also give rise to a symmetric $m_{\nu}$ like the NH case. Thus, the modified $M_D$ to break the symmetry will be given by,
\begin{equation}
M_D=M^{\prime}_D+M_P=
\begin{pmatrix}
a&-b&c+p\\
a&-b+p&c\\
a+p&2b&c\\
\end{pmatrix}.
\end{equation}
Other matrices like $M_{R},M_P,M_{S}^{1},M_{S}^{2},M_{S}^{3}$ will retain their same structure throughout the inverted mass ordering.\\

\section{NUMERICAL ANALYSIS}
\label{sec4}
 The leptonic mixing matrix for active neutrinos depends on three mixing matrices $\theta_{13},\theta_{23}$ and $\theta_{12}$ and one CP-violating phase ($\delta$) for Dirac neutrinos and two Majorana phases $\alpha$ and $\beta$ for Majorana neutrino. Conventionally this Leptonic mass matrix for active neutrino is parameterized as,
 \begin{equation}
 U_{PMNS}={\begin{pmatrix}
 c_{12}c_{13}&s_{12}c_{13}&s_{13}e^{-i\delta}\\
 -s_{12}c_{23}-c_{12}s_{23}s_{13}e^{i\delta}&c_{12}c_{23}-s_{12}s_{23}s_{13}e^{i\delta}&s_{23}c_{13}\\
 s_{12}s_{23}-c_{12}c_{23}s_{13}e^{i\delta}&-c_{12}s_{23}-s_{12}c_{23}s_{13}e^{i\delta}&c_{23}c_{13}\\
 \end{pmatrix}}.P.
 \end{equation}
 The abbreviations used are $c_{ij}=Cos\theta_{ij}$ , $s_{ij}=Sin\theta_{ij}$ and $P$ would be a unit matrix \textbf{1} in the Dirac case but in Majorana case $P=\text{diag}(1,e^{i\alpha},e^{i(\beta+\delta)})$.\\
 The light neutrino mass matrix $M_{\nu}$ is diagonalized by the unitary PMNS matrix as,
 \begin{equation}
\label{eq:2}
 M_{\nu}=U_{PMNS}\ \text{diag}(m_{1},m_{2},m_{3})\ U_{PMNS}^{T},
 \end{equation}
where $m_i$(for $i=1,2,3)$ stands for three active neutrino masses. \\
Since we have included one extra generation of neutrino along with the active neutrinos in our model thus, the final neutrino mixing matrix for the active-sterile mixing takes $4\times4$ form as,
\begin{equation}
V\simeq 
\begin{pmatrix}
(1-\frac{1}{2}RR^{\dagger})U_{PMNS} & R \\ -R^{\dagger}U_{PMNS} & 1-\frac{1}{2}R^{\dagger}R
\end{pmatrix},
\end{equation}
where $R=M_{D}M_{R}^{-1}M_{S}^{T}(M_{S}M_{R}^{-1}M_{S}^{T})^{-1}$ is a $3\times1$ matrix governed by the strength of the active-sterile mixing i.e., the ratio $\frac{\mathcal{O}(M_D)}{\mathcal{O}(M_S)}$.

The sterile neutrino of mass of order eV, can be added to the standard 3-neutrino mass states in NH: $m_1\ll m_2<m_3\ll m_4$ as well as IH: $m_3\ll m_1<m_2\ll m_4$. One can write the diagonal light neutrino mass matrix for NH as $m_{\nu}^{NH}=\text {diag}(0, \sqrt{\Delta m_{21}^{2}}, \sqrt{\Delta m_{21}^{2}+\Delta m_{31}^{2}},\sqrt{\Delta m_{41}^{2}})$ and for IH as,  $m_{\nu}^{IH}=\text{diag}(\sqrt{\Delta m_{31}^{2}},\sqrt{\Delta m_{21}^{2}+\Delta m_{31}^{2}},0,\sqrt{\Delta m_{43}^{2}})$ . The lightest neutrino mass is zero in both the mass ordering as demanded by the MES framework. Here $\Delta m_{41}^{2}(\Delta m_{43}^{2})$ is the active-sterile mass square difference for NH and IH respectively.
As explained in previous section, the non-identical VEV alignment for the Dirac mass matrix in NH and IH produces distinct pattern for the active neutrino mass matrix. The active neutrino mass matrix is obtained using equation \eqref{amass} and the sterile mass is given by equation \eqref{smass}. The complete matrix picture for NH and IH are presented in table \ref{tab:nh} and table \ref{tab:ih} respectively. 

For numerical analysis we have first fixed non-degenerate values for the right-handed neutrino mass parameters as $d=e=10^{13}GeV$ and $f=5\times10^{13}GeV$ so that they can exhibit successful Leptogenesis without effecting the neutrino parameters, which is left for our future study. The mass matrix arises from eq.~\eqref{eq:2} give rise to complex quantities due to the presence of Dirac and the Majorana phases. Since the leptonic CP phases are still unknown, we vary them within their allowed $3\sigma$ ranges (0, 2$\pi$). The Global fit $3\sigma$ values for other parameters like mixing angles, mass square differences are taken from \cite{Capozzi:2016rtj}. One interesting aspect of MES is that if we consider $M_{S}= 
	\begin{pmatrix}g&0&0\\
	\end{pmatrix},$ 
structure, then eventually the parameters from the first column of $M_D$ and $M_R$ matrices goes away and does not appear in the light neutrino mass matrix given by \eqref{amass}. The same argument justify the disappearance of the model parameters in the other two cases also. Hence the active neutrino mass matrix emerging from our model matrices is left with three parameters for each case. Comparing the model mass matrix with the one produced by light neutrino parameters given by eq. \eqref{eq:2}, we numerically evaluate the model parameters satisfying the current bound for the neutrino parameters and establish correlation among various model and oscillation parameters within $3\sigma$ bound \footnote{The evaluated model parameters are complex in nature as they are solved using the complex matrix. While plotting, we have used the absolute values for the parameters.}. Three assessment for each distinct structures of $M_D$ for both normal and inverted hierarchy cases are carried out in the following subsections. 
\subsection{NORMAL HIERARCHY}\label{ss1}
\begin{figure}
\includegraphics[width=8cm,height=5.3cm]{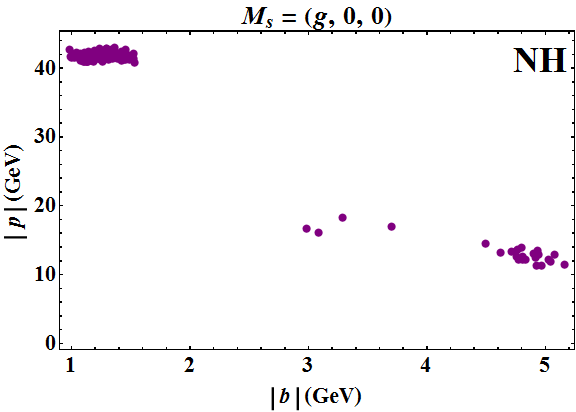}
\includegraphics[width=8cm,height=5.3cm]{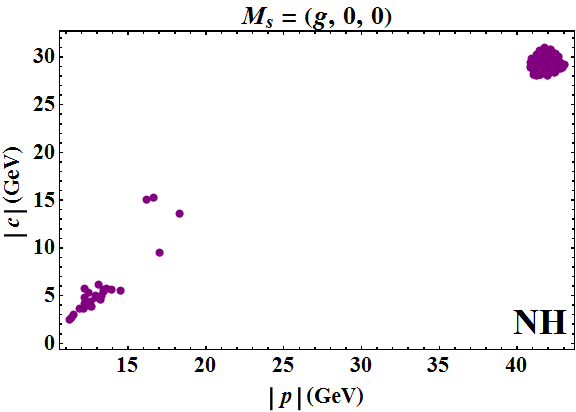}\\
\includegraphics[width=8cm,height=5.3cm]{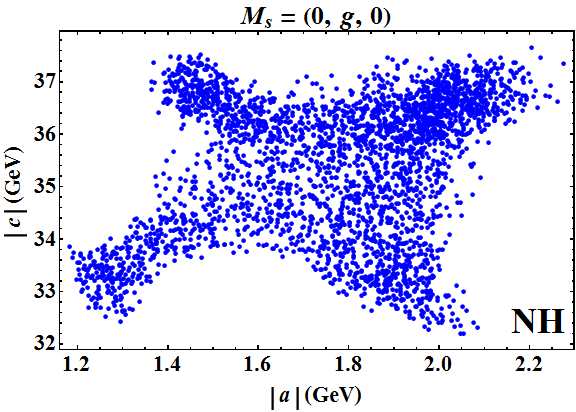}
\includegraphics[width=8cm,height=5.3cm]{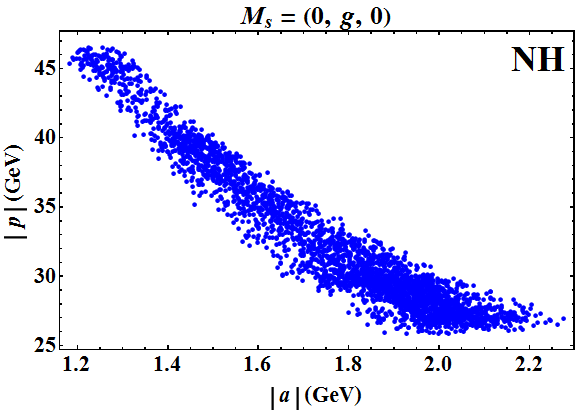}\\
\includegraphics[width=8cm,height=5.3cm]{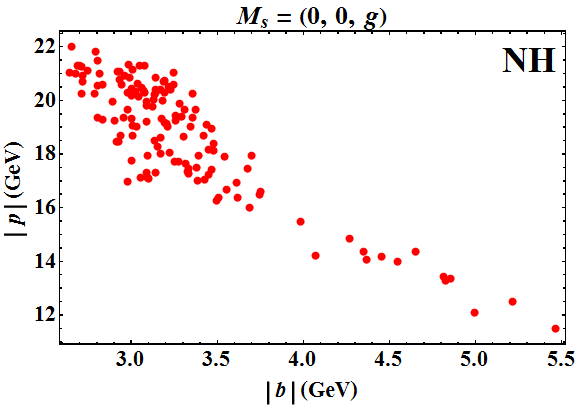}
\includegraphics[width=8cm,height=5.3cm]{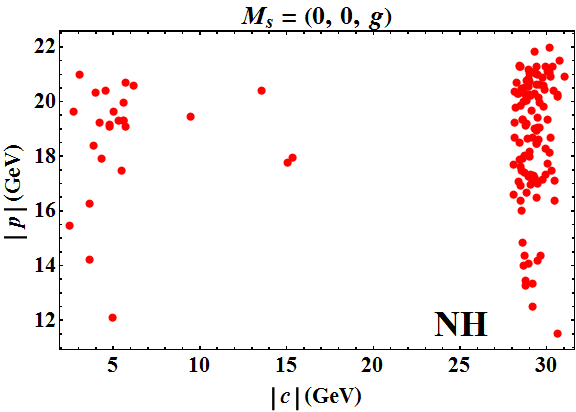}\\

\caption{Variation of model parameters among themselves for the NH pattern.}\label{modelnh}
\end{figure}

\begin{table}
\begin{center}
\begin{tabular}{|c|c|c|}
\hline
Mass ordering & Structures & $m_{\nu}$ \\
\hline
NH(Case-I) & 
$\begin{aligned}
&
M_R=\begin{pmatrix}
d&0&0\\
0&e&0\\
0&0&f\\
\end{pmatrix}\\
& M_{D}= \begin{pmatrix}
a&b&c+p\\
a&b+p&c\\
a+p&b&c\\
\end{pmatrix}\\
& M_{S}^{1}= \begin{pmatrix}
g&0&0\\
\end{pmatrix}\\
\end{aligned}$
& $m_{\nu}= -\begin{pmatrix}
\frac {b^2} {e} + \frac {(c + p)^2} {f} &\frac {b (b + 
	p)} {e} + \frac {c (c + 
	p)} {f} &\frac {b^2} {e} + \frac {c (c + p)} {f} \\
\frac {b (b + p)} {e} + \frac {c (c + p)} {f} &\frac {(b + 
	p)^2} {e} + \frac {c^2} {f} &\frac {b (b + 
	p)} {e} + \frac {c^2} {f} \\
\frac {b^2} {e} + \frac {c (c + p)} {f} &\frac {b (b + 
	p)} {e} + \frac {c^2} {f} &\frac {b^2} {e} + \frac {c^2}
{f} \\
\end{pmatrix}$ \\
\hline
NH(Case-II) & 
$\begin{aligned}
&
M_R=\begin{pmatrix}
d&0&0\\
0&e&0\\
0&0&f\\
\end{pmatrix}\\
& M_{D}= \begin{pmatrix}
a&b&c+p\\
a&b+p&c\\
a+p&b&c\\
\end{pmatrix}\\
& M_{S}^{2}= \begin{pmatrix}
0&g&0\\
\end{pmatrix}\\
\end{aligned}$
& $m_{\nu}= -\begin{pmatrix}
\frac {a^2} {d} + \frac {(c + 
	p)^2} {f} &\frac {a^2} {d} + \frac {c (c + 
	p)} {f} &\frac {a (a + p)} {d} + \frac {c (c + p)} {f} \\
\frac {a^2} {d} + \frac {c (c + 
	p)} {f} &\frac {a^2} {d} + \frac {c^2} {f} &\frac {a (a \
	+ p)} {d} + \frac {c^2} {f} \\
\frac {a (a + p)} {d} + \frac {c (c + p)} {f} &\frac {a (a + 
	p)} {d} + \frac {c^2} {f} &\frac {(a + 
	p)^2} {d} + \frac {c^2} {f} \\
\end{pmatrix}$ \\
\hline
NH(Case-III) & 
$\begin{aligned}
&
M_R=\begin{pmatrix}
d&0&0\\
0&e&0\\
0&0&f\\
\end{pmatrix}\\
& M_{D}= \begin{pmatrix}
a&b&c+p\\
a&b+p&c\\
a+p&b&c\\
\end{pmatrix}\\
& M_{S}^{3}= \begin{pmatrix}
0&0&g\\
\end{pmatrix}\\
\end{aligned}$
& $m_{\nu}= -\begin{pmatrix}\frac {a^2} {d} + \frac {b^2} {e} &\frac {a^2} {d} + \frac {b (b + 
	p)} {e} &\frac {a (a + p)} {d} + \frac {b^2} {e} \\
\frac {a^2} {d} + \frac {b (b + p)} {e} &\frac {a^2} {d} + \frac {(b +
	p)^2} {e} &\frac {a (a + p)} {d} + \frac {b (b + p)} {e} \\
\frac {a (a + p)} {d} + \frac {b^2} {e} &\frac {a (a + 
	p)} {d} + \frac {b (b + p)} {e} &\frac {(a + 
	p)^2} {d} + \frac {b^2} {e} \\
\end{pmatrix}$ \\
\hline
\end{tabular}
\caption{The light neutrino mass matrices and the corresponding $M_D$ and $M_R$ matrices for three different structures of $M_S$ under NH pattern. }\label{tab:nh}
\end{center}
\end{table}
\begin{table}
	\begin{tabular}{|c|c|c|c|}
		\hline
		Case&$M_S$&$m_s(eV)$&$R$\\
		\hline
		\hline
		I& $M_{S}^{1}= \begin{pmatrix}
		g&0&0\\
		\end{pmatrix}$
		&$m_s\simeq\frac{g^2}{10^4}$&$R^T\simeq{\begin{pmatrix}
			\frac{a}{g}& \frac{a}{g}& \frac{a+p}{g}\\
			\end{pmatrix}}^T$\\
		\hline
		II& $M_{S}^{2}= \begin{pmatrix}
		0&g&0\\
		\end{pmatrix}$&$m_s\simeq\frac{g^2}{10^4}$&$R^T\simeq{\begin{pmatrix}
			\frac{b}{g}& \frac{b+p}{g}& \frac{b}{g}\\
			\end{pmatrix}}^T$\\
		\hline
		III& $M_{S}^{3}= \begin{pmatrix}
		0&0&g\\
		\end{pmatrix}$&$m_s\simeq\frac{g^2}{5\times10^4}$&$R^T\simeq{\begin{pmatrix}
			\frac{c+p}{g}& \frac{c}{g}& \frac{c}{g}\\
			\end{pmatrix}}^T$\\
		\hline
	\end{tabular}
	\caption{Sterile neutrino mass and active-sterile mixing matrix for three different $M_S$ structures under NH pattern.}\label{tab:msnh}
\end{table}
For the diagonal charged lepton mass we have chosen a non-trivial VEV alignment resulting a specific pattern in Dirac mass hence a broken $\mu-\tau$ symmetry along with non-zero reactor mixing angle is achieved. The complete picture for  active neutrino mass matrices and the sterile sector for different cases are shown in table \ref{tab:nh} and \ref{tab:msnh} respectively. For each $M_S$ structure, three variables are there in the light neutrino mass matrix. After solving them by comparing with the light neutrino mass, we obtain some correlation plots which redefines our model parameters with more specific bounds. Correlation among various model parameters in NH are shown in fig. \ref{modelnh}. One would notice the fact in the active mass matrices from table \ref{tab:nh} and \ref{tab:ih} is that, in the limit $p\rightarrow0$ and all the model parameters become equal to one, the matrix takes the form of a democratic mass matrix. Hence the $p$ parameters brings out a phenomenological change  and plays an important role in our study. Various plots with the model parameter $p$ are shown below in fig. \ref{psnh} and \ref{psih}.


\begin{figure}
\includegraphics[width=8cm,height=5.3cm]{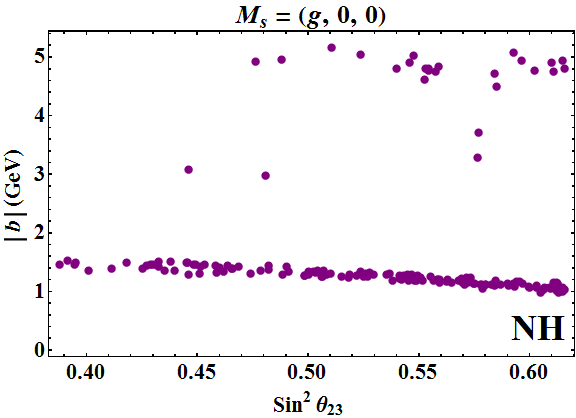}
\includegraphics[width=8cm,height=5.3cm]{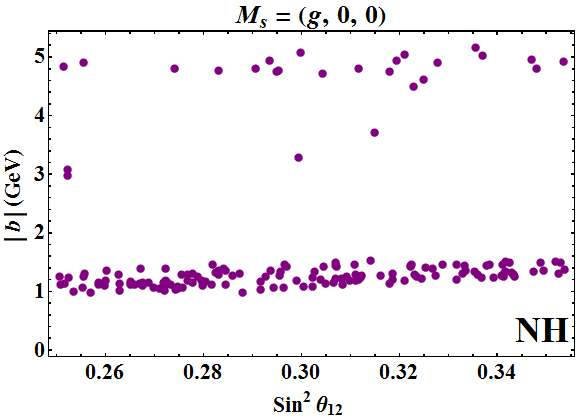}\\
\includegraphics[width=8cm,height=5.3cm]{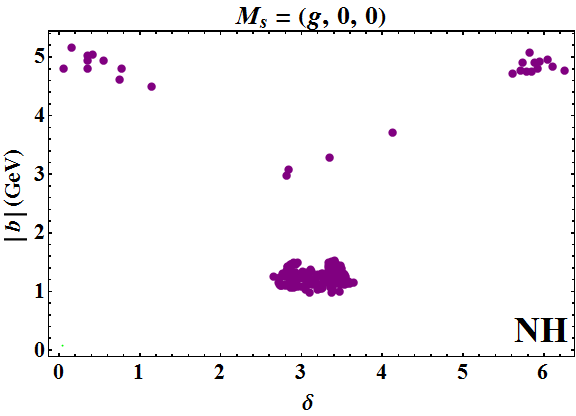}
\includegraphics[width=8cm,height=5.3cm]{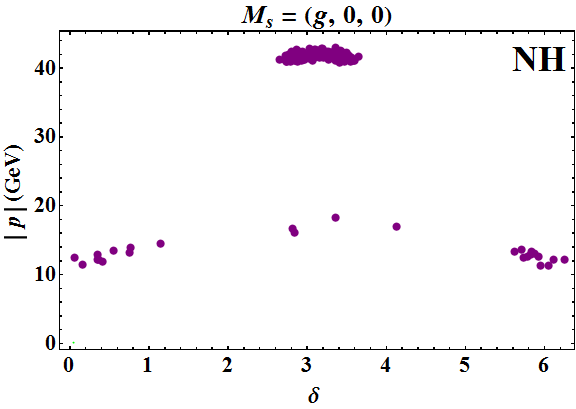}\\
\includegraphics[width=8cm,height=5.3cm]{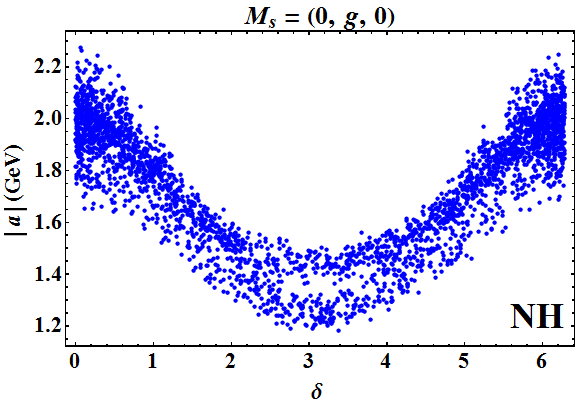}
\includegraphics[width=8cm,height=5.3cm]{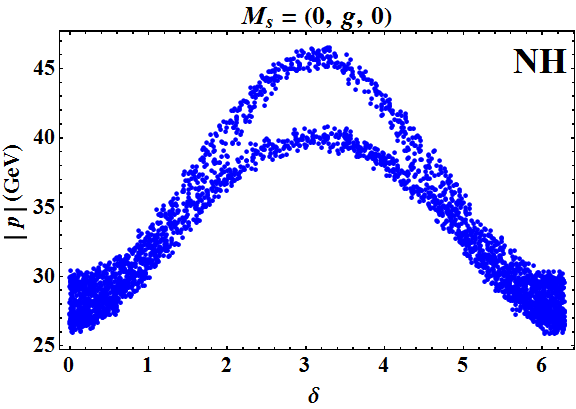}\\
\includegraphics[width=8cm,height=5.3cm]{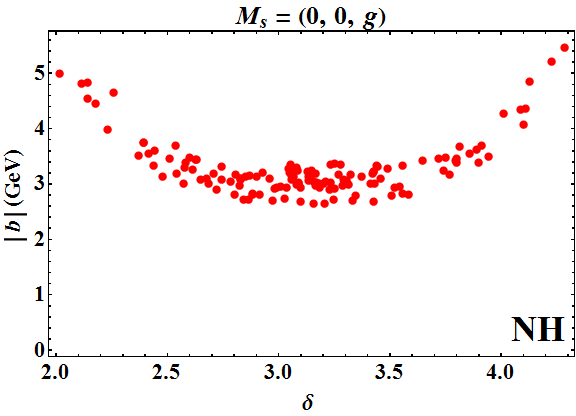}
\includegraphics[width=8cm,height=5.3cm]{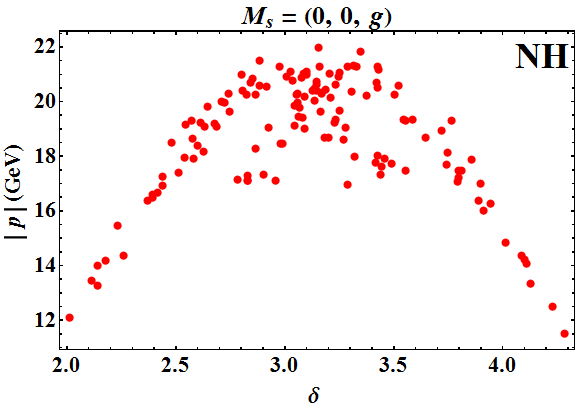}

\caption{Correlation plots among various model parameters and light neutrino parameters(within $3\sigma$ bound) in NH. The Dirac CP phase shows a good correlation with the model parameters than the other light neutrino parameters.}\label{2nh}
\end{figure}
\begin{figure}
	\includegraphics[scale=.32]{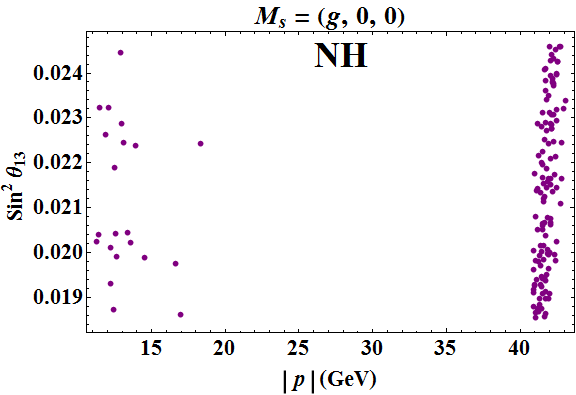}
	\includegraphics[scale=.32]{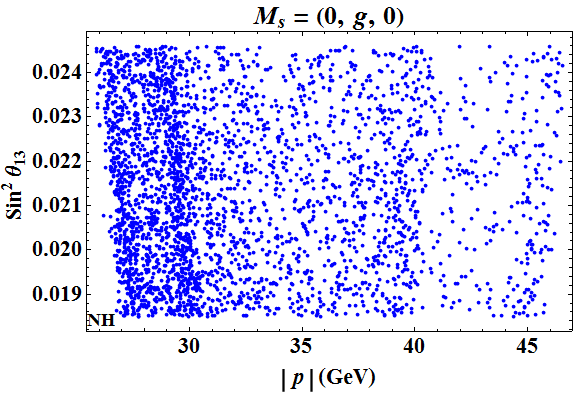}
	\includegraphics[scale=.32]{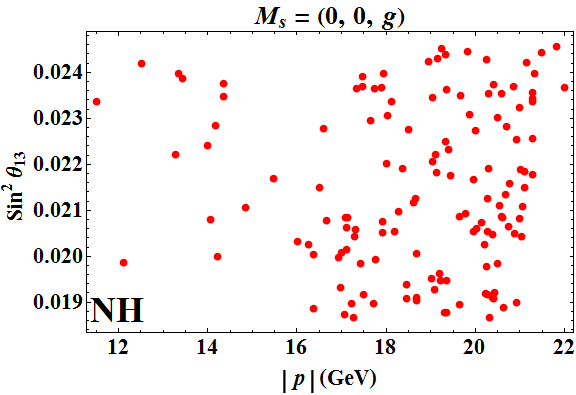}
	\caption{Variation of $Sine$ of reactor mixing angle with $p$, which is responsible for the generation of reactor mixing angle($\theta_{13}$) for NH.}\label{psnh}
\end{figure}
\begin{figure}
\includegraphics[scale=.32]{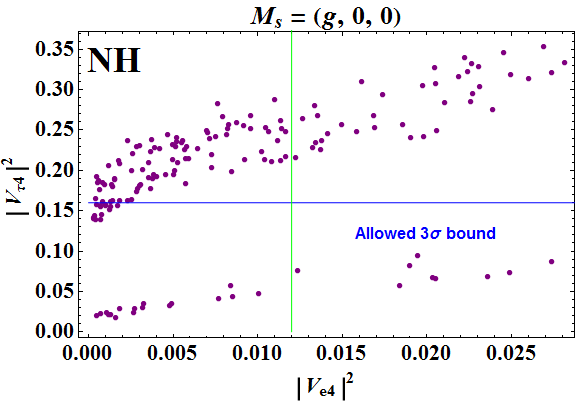}
\includegraphics[scale=.32]{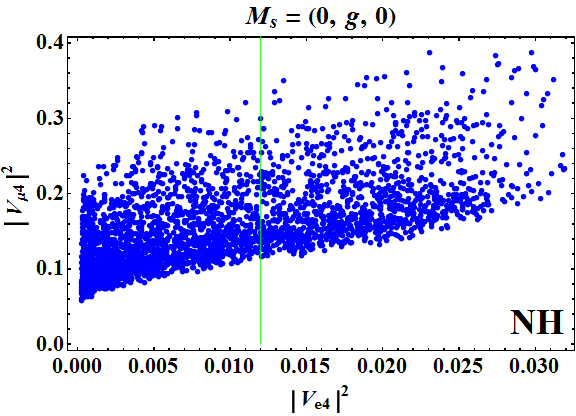}
\includegraphics[scale=.32]{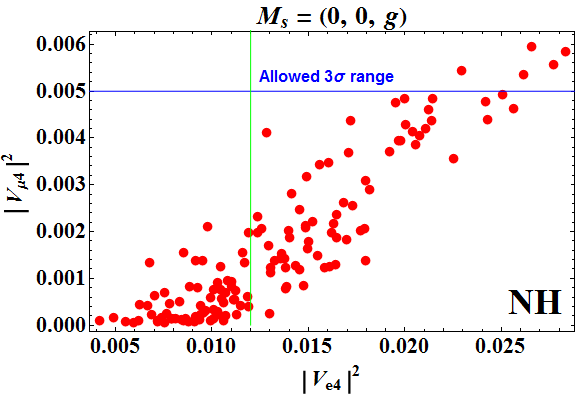}
\caption{Allowed bound for the active-sterile mixing matrix elements in NH. The green line  is the lower bound for $|V_{e4}|^2$ and the blue line in the first plot gives the upper bound for $|V_{\tau4}|^2$ while in the third plot it gives the lower bound for $|V_{\mu4}|^2$}\label{4nh}
\end{figure}
As $m_s$ depends only on $M_R$ and $M_S$, so due to the non-degenerate value of $M_R$, the $m_s$ structure let us study the active-sterile mixing strength $R$. The active-sterile mixing matrix also have a specific form due to the particular $M_S$ structure.
\pagebreak
\subsection{INVERTED HIERARCHY}\label{ss2}
\begin{table}
\begin{center}
\begin{tabular}{|c|c|c|}
\hline
Mass ordering & Structures & $m_{\nu}$ \\
\hline
IH
(Case-I)& 
$\begin{aligned}
&
M_R=\begin{pmatrix}
d&0&0\\
0&e&0\\
0&0&f\\
\end{pmatrix}\\
& M_{D}= \begin{pmatrix}
a&-b&c+p\\
a&-b+p&c\\
a+p&2b&c\\
\end{pmatrix}\\
& M_{S}^{1}= \begin{pmatrix}
g&0&0\\
\end{pmatrix}\\
\end{aligned}$
& $m_{\nu}= -\begin{pmatrix}
\frac {b^2} {e} + \frac {(c + p)^2} {f} &  \frac {b(b - 
	p)} {e} + \frac {c (c + 
	p)} {f} & \frac {-2 b^2} {e} + \frac {c (c + p)} {f} \\
\frac {b(b - p)} {e} + \frac {c (c + p)} {f} &\frac {(b - 
	p)^2} {e} + \frac {c^2} {f} &\frac {-2 b (b - 
	p)} {e} + \frac {c^2} {f} \\
-\frac {2 b^2} {e} + \frac {c (c + p)} {f} & - \frac {2 b (b - 
	p)} {e} + \frac {c^2} {f} &\frac {4 b^2} {e} + \frac {c^2}
{f} \\
\end{pmatrix}$ \\
\hline
IH(Case-II) & 
$\begin{aligned}
&
M_R=\begin{pmatrix}
d&0&0\\
0&e&0\\
0&0&f\\
\end{pmatrix}\\
& M_{D}= \begin{pmatrix}
a&-b&c+p\\
a&-b+p&c\\
a+p&2b&c\\
\end{pmatrix}\\
& M_{S}^{2}= \begin{pmatrix}
0&g&0\\
\end{pmatrix}\\
\end{aligned}$
& $m_{\nu}= -\begin{pmatrix}
\frac {a^2} {d} + \frac {(c + 
	p)^2} {f} &\frac {a^2} {d} + \frac {c (c + 
	p)} {f} &\frac {a (a + p)} {d} + \frac {c (c + p)} {f} \\
\frac {a^2} {d} + \frac {c (c + 
	p)} {f} &\frac {a^2} {d} + \frac {c^2} {f} &\frac {a (a \
	+ p)} {d} + \frac {c^2} {f} \\
\frac {a (a + p)} {d} + \frac {c (c + p)} {f} &\frac {a (a + 
	p)} {d} + \frac {c^2} {f} &\frac {(a + 
	p)^2} {d} + \frac {c^2} {f} \\
\end{pmatrix}$ \\
\hline
IH(Case-III) & 
$\begin{aligned}
&
M_R=\begin{pmatrix}
d&0&0\\
0&e&0\\
0&0&f\\
\end{pmatrix}\\
& M_{D}= \begin{pmatrix}
a&-b&c+p\\
a&-b+p&c\\
a+p&2b&c\\
\end{pmatrix}\\
& M_{S}^{3}= \begin{pmatrix}
0&0&g\\
\end{pmatrix}\\
\end{aligned}$
& $m_{\nu}= -\begin{pmatrix}\frac {a^2} {d} + \frac {(b^2} {e} &\frac {a^2} {d} - \frac {b (-b + 
	p)} {e} &\frac {a (a + p)} {d} - \frac {2 b^2} {e} \\
\frac {a^2} {d} - \frac {b (-b + 
	p)} {e} &\frac {a^2} {d} + \frac {(b - 
	p)^2} {e} &\frac {a (a + p)} {d} + \frac {2 b (b - p)} {e} \\
\frac {a (a + p)} {d} + \frac {-2 b^2} {e} &\frac {a (a + 
	p)} {d} - \frac {2 b (b - p)} {e} &\frac {(a + 
	p)^2} {d} + \frac {4 b^2} {e} \\
\end{pmatrix}$ \\
\hline
\end{tabular}
\end{center}
\caption{The light neutrino mass matrices and the corresponding $M_D$ and $M_R$ matrices for three different structures of $M_S$ under IH pattern.}\label{tab:ih}
\end{table}
In this section we will discuss the inverted mass ordering ($i.e., m_2>m_1>m_3$) of the neutrinos. Referring to \cite{Zhang:2011vh}, we have introduced a new flavon as, $\langle\varphi^{\prime}\rangle=(2v,-v,-v)$ in the Yukawa Lagrangian for the Dirac mass term, so that this model can exhibit inverted hierarchy. A detailed discussion has already been carried out in previous section \ref{sec3}. 

Numerical procedure for IH is analogous to the NH. Here also we have considered three distinguished cases for $M_S$, which is responsible for three separate $m_\nu$ matrices. A brief picture for the matrices has shown in table \ref{tab:ih}.

In table \ref{tab:ih1}, three different $M_S$ structures are shown, which lead to various $m_s$ and $R$ values. Unlike the normal ordering, a deviation from the common track is observed in $R$ matrix for the second case ($M_S^{2}=(0,g,0)$). This occurs due to the change in $M_D$ matrix structure for non-identical VEV alignment.
\begin{center}
\begin{table}
\begin{tabular}{|c|c|c|c|}
\hline
Case&$M_S$&$m_s(eV)$&$R$\\
\hline
\hline
I& $M_{S}^{1}= \begin{pmatrix}
g&0&0\\
\end{pmatrix}$
&$m_s\simeq\frac{g^2}{10^4}$&$R^T\simeq{\begin{pmatrix}
\frac{a}{g}& \frac{a}{g}& \frac{a+p}{g}\\
\end{pmatrix}}^T$\\
\hline
II& $M_{S}^{2}= \begin{pmatrix}
0&g&0\\
\end{pmatrix}$&$m_s\simeq\frac{g^2}{10^4}$&$R^T\simeq{\begin{pmatrix}
\frac{-b}{g}& \frac{-b+p}{g}& \frac{2b}{g}\\
\end{pmatrix}}^T$\\
\hline
III& $M_{S}^{3}= \begin{pmatrix}
0&0&g\\
\end{pmatrix}$&$m_s\simeq\frac{g^2}{5\times10^4}$&$R^T\simeq{\begin{pmatrix}
\frac{c+p}{g}& \frac{c}{g}& \frac{c}{g}\\
\end{pmatrix}}^T$\\
\hline
\end{tabular}
\caption{Sterile neutrino mass and active-sterile mixing matrix for three different $M_S$ structures under IH pattern. }\label{tab:ih1}
\end{table}
\end{center}
\begin{figure}
\includegraphics[width=8cm,height=5.3cm]{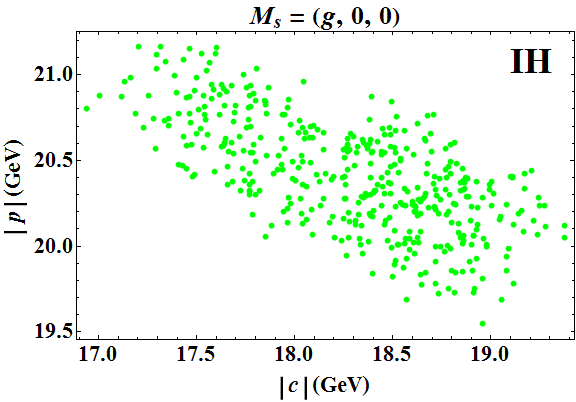}
\includegraphics[width=8cm,height=5.3cm]{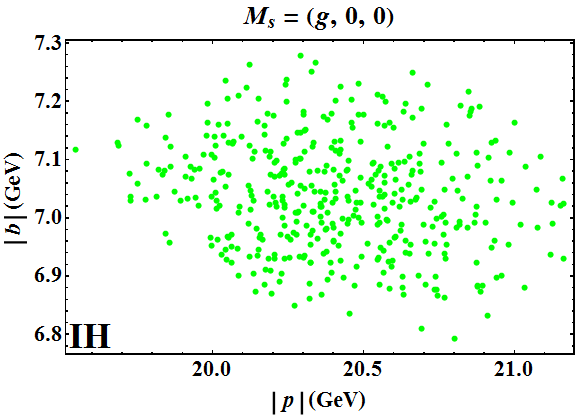}
\includegraphics[width=8cm,height=5.3cm]{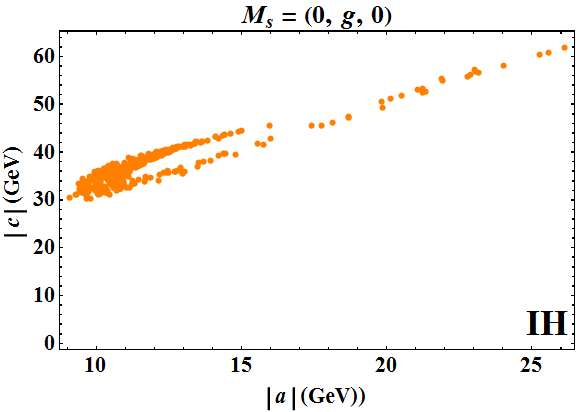}
\includegraphics[width=8cm,height=5.3cm]{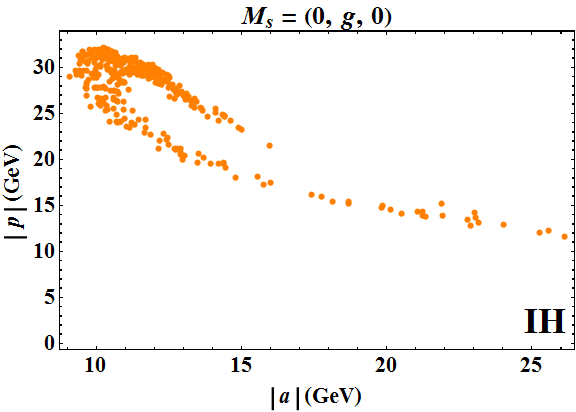}
\includegraphics[width=8cm,height=5.3cm]{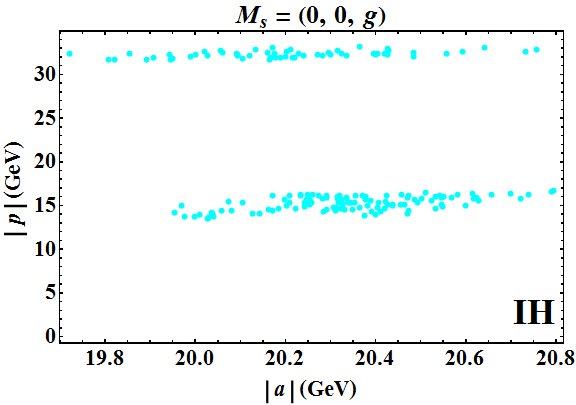}
\includegraphics[width=8cm,height=5.3cm]{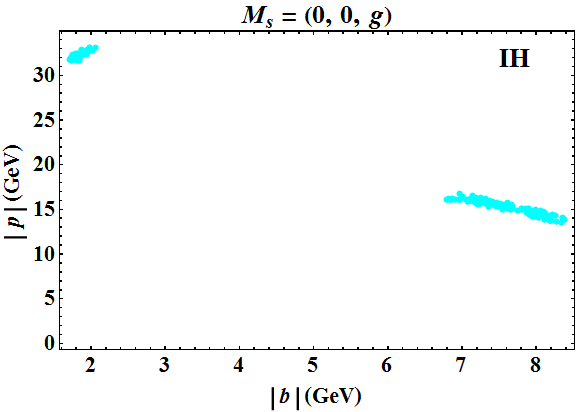}
\caption{Constrained region of model parameters in case of IH pattern.}\label{modelih}
\end{figure}
\begin{figure}
\includegraphics[width=8cm,height=5.3cm]{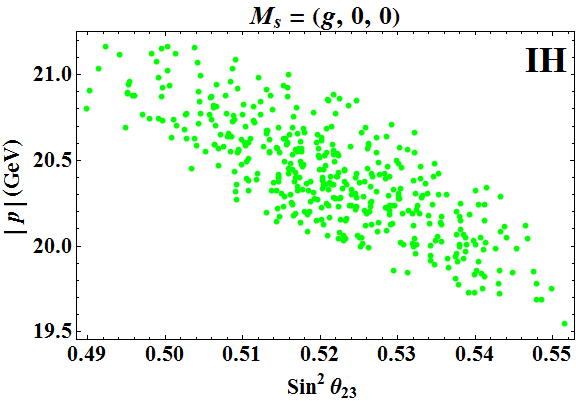}
\includegraphics[width=8cm,height=5.3cm]{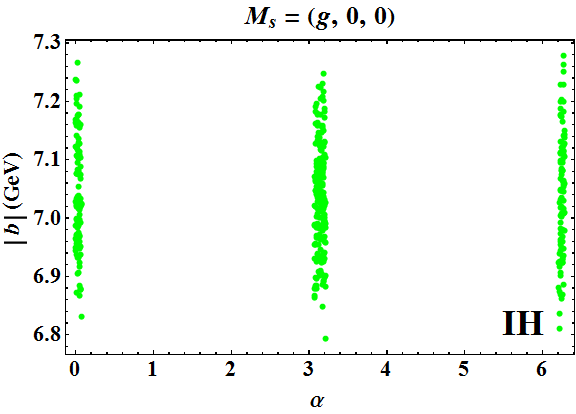}
\includegraphics[width=8cm,height=5.3cm]{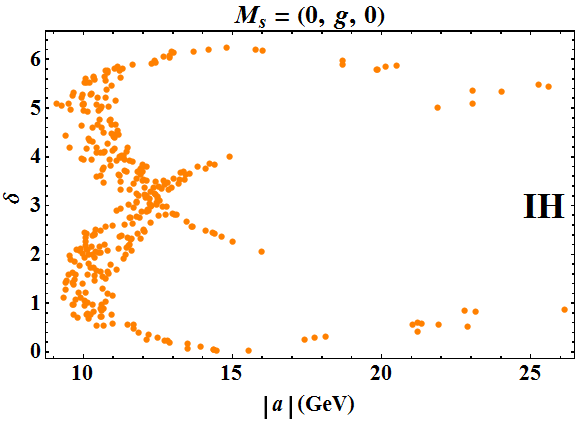}
\includegraphics[width=8cm,height=5.3cm]{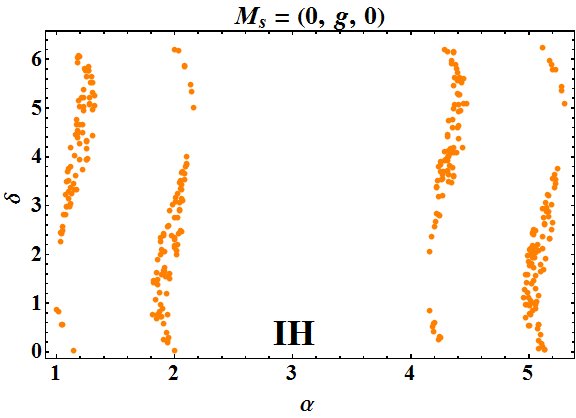}
\includegraphics[width=8cm,height=5.3cm]{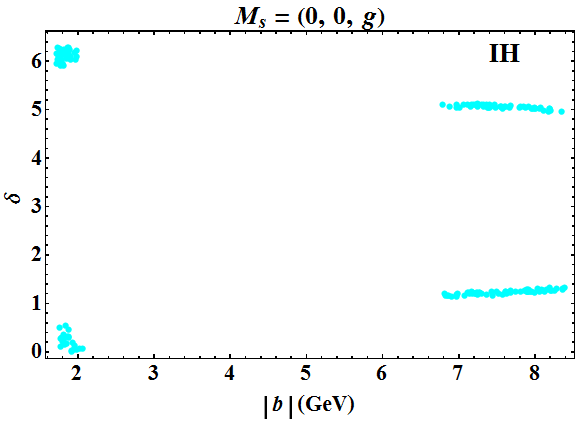}
\includegraphics[width=8cm,height=5.3cm]{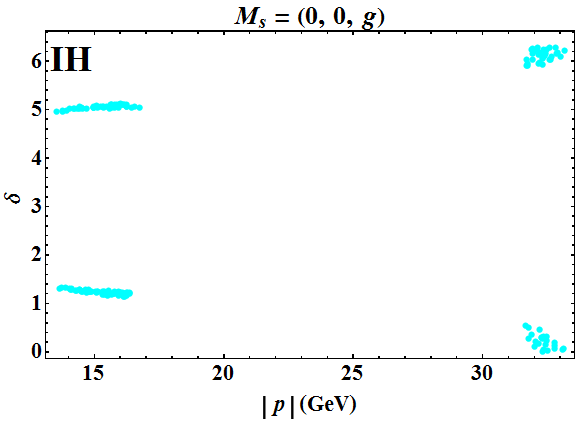}
\includegraphics[width=8cm,height=5.3cm]{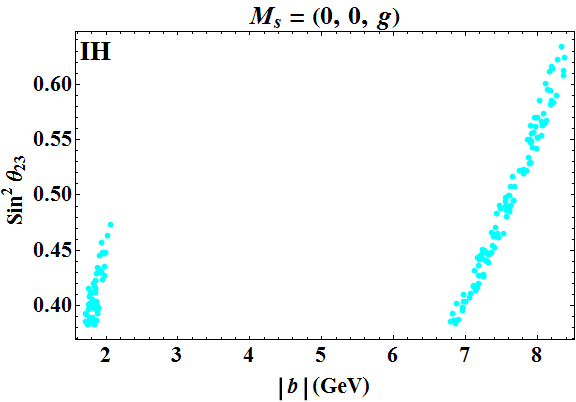}

\caption{Correlation plots among various model parameters with light neutrino parameters in IH pattern.}\label{2ih}
\end{figure}
\begin{figure}
	\includegraphics[scale=.30]{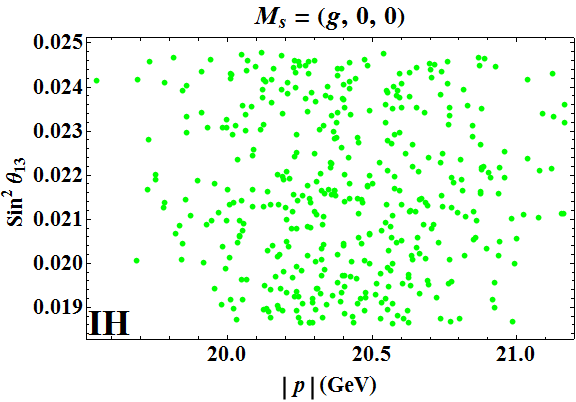}
	\includegraphics[scale=.30]{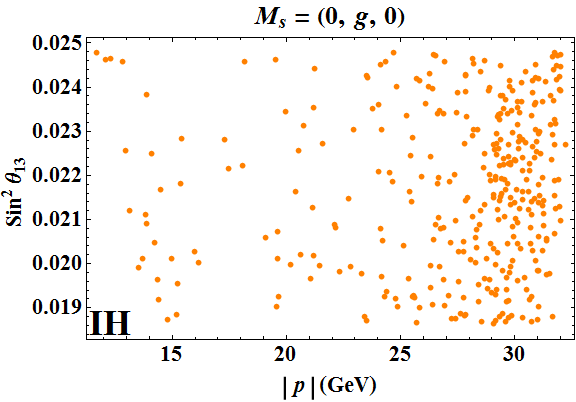}
	\includegraphics[scale=.30]{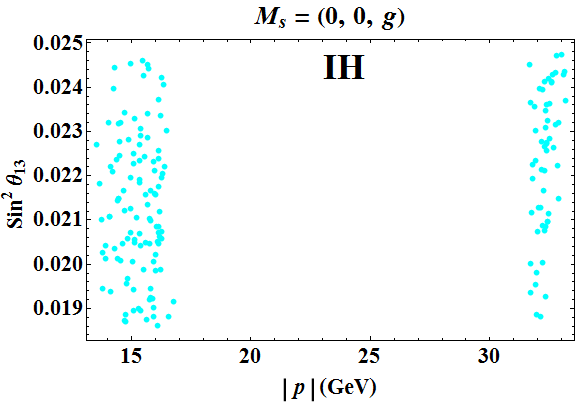}
	\caption{Variation of $p$ with the $Sine$ of reactor mixing angle for IH. The third structure of $M_S$ shows a constrained region for the model parameter.}\label{psih}
\end{figure}
\begin{figure}
\includegraphics[scale=.32]{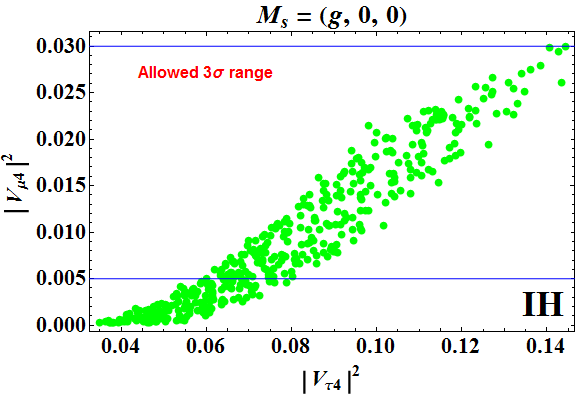}
\includegraphics[scale=.32]{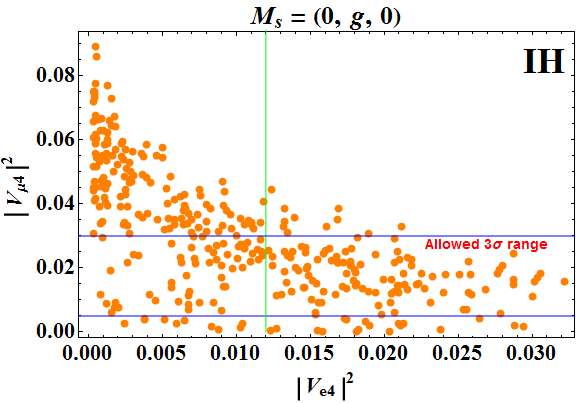}
\includegraphics[scale=.32]{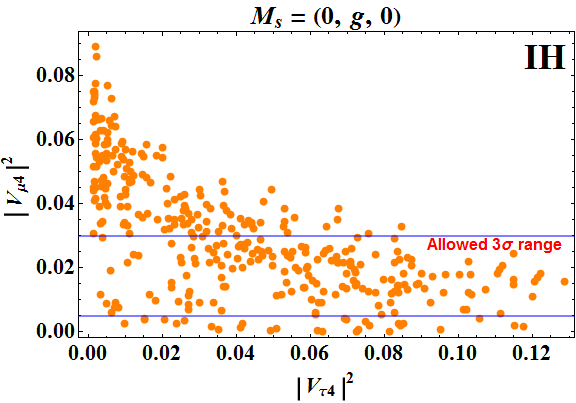}
\includegraphics[scale=.32]{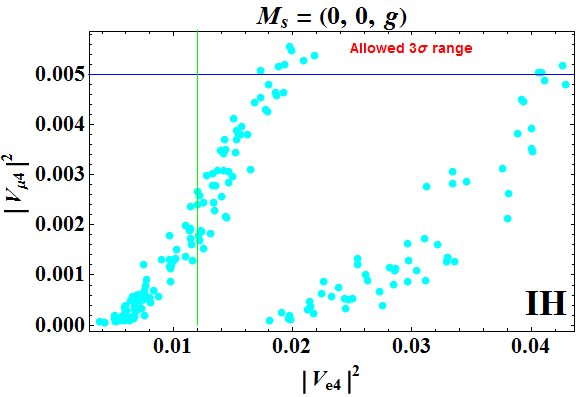}
\caption{Allowed bound for active-sterile mixing matrix elements in IH. The blue solid line gives the upper and lower bound for $|V_{\mu4}|^2$ along the y-axix while solid green line gives the lower bound for $|V_{e4}|^2$ along the x-axis. }\label{4ih}
\end{figure}
\begin{figure}
	\includegraphics[scale=.35]{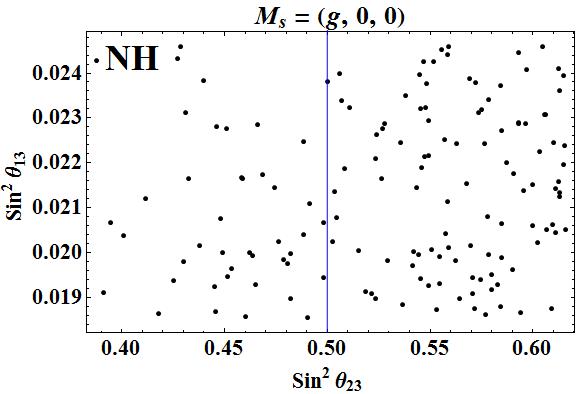}
	\includegraphics[scale=.35]{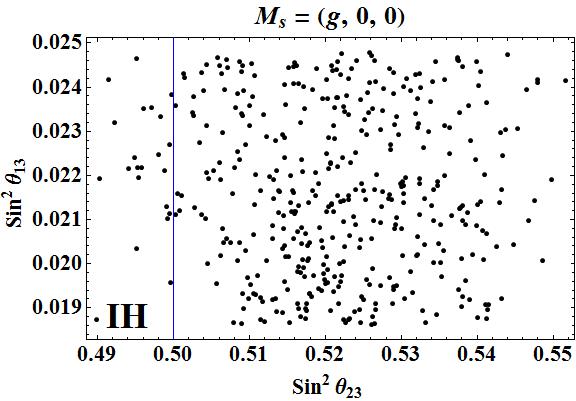}
	\includegraphics[scale=.35]{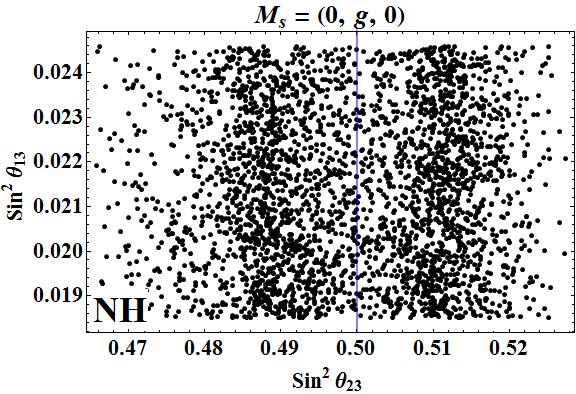}
	\includegraphics[scale=.35]{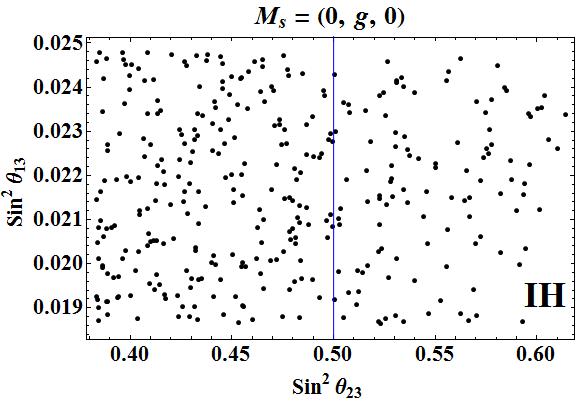}
	\includegraphics[scale=.35]{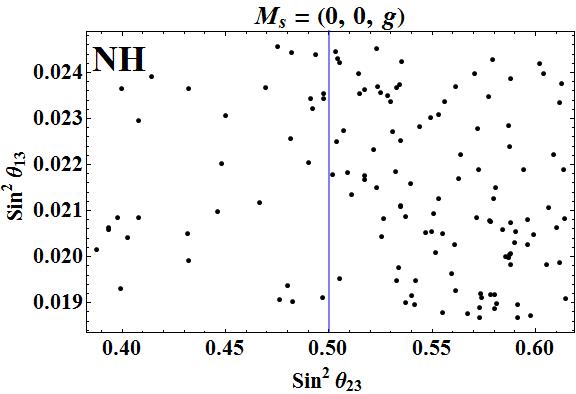}
	\includegraphics[scale=.35]{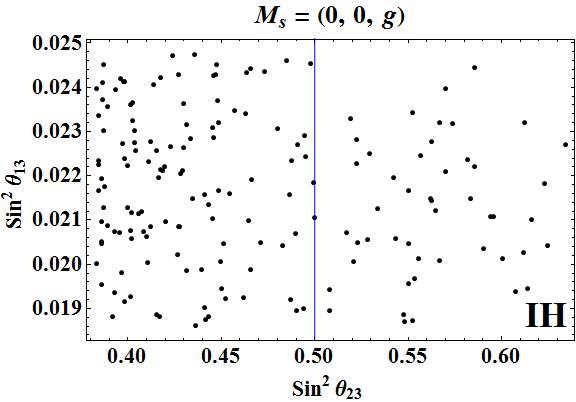}
	\caption{Variation of $Sin^2\theta_{13}$ vs. $Sin^2\theta_{23}$ in both the mass ordering for all the three $M_S$ structures. }\label{1323}
\end{figure}
\clearpage




\section{SUMMARY AND CONCLUSION}
\label{sec5}
In this paper we have investigated the extension of low scale SM type-I seesaw $i.e.,$ the minimal extended seesaw, which restricts active neutrino masses to be within sub-eV scale and generates an eV scale light sterile neutrino. $A_4$ based flavor model is extensively studied along with a discrete Abelian symmetry $Z_4$ and $Z_3$ to construct the desired Yukawa coupling matrices. Under this MES framework the Dirac mass $M_D$ is a $3\times3$ complex matrix. The Majorana mass matrix $M_R$, which arises due to the coupling of right-handed neutrinos with the anti-neutrinos is also a $3\times 3$ complex symmetric diagonal matrix with non-degenerate eigenvalues. A singlet $S_i$ (where $i=1,2,3$) is considered which couples with the right-handed neutrinos ($\nu_{Ri};i=1,2,3$) and produces a singled row  $1\times 3$ $ M_S$ matrix with one non-zero entry. In earlier studies like \cite{Zhang:2011vh,Nath:2016mts}, the $A_4$ flavor symmetry in MES was implemented with a little description. In our phenomenological study, we have addressed non-zero reactor mixing angle with a detailed discussion on VEV alignment of the flavon fields, which were discussed under the light of flavor symmetry within this MES framework. Three separate cases are carried out for both NH and IH for three $M_S$ structures. Within the active neutrino mass matrix, the common $\mu-\tau$ symmetry is broken along with $\theta_{13}\neq 0$ by adding a new matrix ($M_P$) to the Dirac mass matrix. 

Both normal and inverted cases are analyzed independently for three $M_S$ structures in this work. We have used similar numerical techniques for solving model parameters in both the cases (NH \& IH) and plotted them among themselves as well as with the light neutrino parameters. The plots in fig. \ref{modelnh} , \ref{2nh}, \ref{modelih}, \ref{2ih} show constrained parameter space in the active neutrino sector in case of NH and IH for various $M_S$ structure. In most of the cases the parameter space is narrow, which can be verified or falsified in future experiments. In the $m_{\nu}$ matrix, the $\mu-\tau$ symmetry is broken due to the extra term added to the Dirac mass matrix. The variation of $Sin^2\theta_{13}$ with $p$ plotted in fig. \ref{psnh} and \ref{psih} for NH \& IH respectively. Within NH, the first structure of $M_S$ shows a better constrained region for the model parameter ($p$) than the other two structures. Whereas in IH case, the third $M_S$ structure gives a relatively narrower region than that obtained for the other two structures. In fig \ref{1323} we have plotted {\it sine} squared of the reactor mixing angle against the atmospheric mixing angle. Within the NH mode $M_S^1$ and $M_S^3$ structures favors the upper octant of $Sin^2\theta_{23}$ accommodating more numbers of data points whereas within IH mode, $M_S^2$ and $M_S^3$ structures are more favorable in the lower octant. The $M_S^1$ structure in IH mode is heavily constrained within the upper octant of the atmospheric mixing angle. On the other hand the second structure of $M_S$ in the NH case is showing deviation from the maximal atmospheric mixing, having dense regions in either octant hence not constrained for $Sin^2\theta_{23}$.\\
The active-sterile mixing phenomenology is also carried out under the same MES framework. The fourth column of the active-sterile mixing matrix is generated and solved the elements with an acceptable choice of Yukawa coupling. Apart from generating non-zero $\theta_{13}$, the matrix element of $M_P$ has an important role to play in the active-sterile mixing. As we can see in table \ref{tab:msnh} and \ref{tab:ih1}, $p$ has an active participation in differentiating the elements of $R$ matrix. We have plotted the mixing matrix elements ($V_{e4},V_{\mu4},V_{\tau4}$) within themselves as shown in fig. \ref{4nh} and \ref{4ih}. The SK collaboration limits $|V_{\mu 4}|^2 < 0.04$ for $\Delta m_{41}^2>0.1eV^2$ at 90\% CL by considering $|V_{e4}|^2=0$ \cite{Abe:2014gda}. For $\Delta m_{41}^2\sim 1eV^2$, the IceCube DeepCore collaboration suggested that $|V_{\mu 4}|^2 < 0.03$ with $|V_{\tau 4}|^2<0.15$ and $|V_{e4}|^2$ is around 0.012 at 90\% CL \cite{Aartsen:2017bap,Thakore:2018lgn}. In particular,for various mass range of $\Delta m_{41}^2$ there are more fascinating results about active-sterile mixing however this is beyond the scope of this paper. These bounds are consistent with some of our model structures. In NH case, the first and the third $M_S$ structure show an allowed $3\sigma$ range for the mixing parameters but no such mutual allowed range is obtained for the second structure of $M_S$. The plots in fig. \ref{4ih} \ shows the IH case for the mixing elements. The first structure of $M_S$ covers a wider range of allowed data points within the $3\sigma$ bound than the other two case.
 We have not shown any plots relating active-sterile mass squared difference with the active mass and active-sterile mixing elements, because from table VI and VIII one can see that the sterile mass is emerging as $g^2/10^4$, $g^2/10^4$ and $g^2/{5\times10^4}$ for respective $M_S^1$, $M_S^2$ and $M_S^3$ structures and the active sterile mixing matrix also contains $g$ exclusively. Since $g$ is evaluated using active-sterile mass squared difference so a comparison of this mass squared difference with the active-sterile mixing matrix would be needless within this numerical approach. Since we are focusing on model building aspects and here we are taking bounds for the light neutrino parameters from global fit data to verify/predict our model. Plot between any two light neutrino observable would be simply a presentation between two global fit data sets which does not carry any significance in our study. This argument also support why we haven't shown any plot between between active-sterile mass squared differences/angles with the active neutrino masses/angles Moreover the active-sterile mixing matrix elements and mixing angles are more or less represent the same phenomenology but we prefer to show correlation among the matrix elements over the mixing angles as the bounds for the mixing elements are more auspicious than the angles (although the mixing elements depends on the mixing angles).\\
 Authors in \cite{Barry:2011wb, Zhang:2011vh} have discussed active and sterile phenomenology by considering the same MES framework under $A_4$ flavor symmetry. The light neutrino mass matrix ($m_{\nu}$) is diagonalized using the tri-bimaximal mixing matrix and it is found to be $\mu-\tau$ symmetric matrix. On the other hand, as per current experimental demand \cite{An:2012eh}, in our current work, we have addressed non-zero reactor mixing angle by adding a correction term in the $M_D$, which break the trivial $\mu-\tau$ symmetry in $m_{\nu}$, that was considered zero in earlier studies. The extra correction term (that is added to $M_D$) has substantial influence to the reactor mixing angle, which is discussed in the last paragraph. In addition to their previous work \cite{Barry:2011wb, Zhang:2011vh}, we have constructed our model with new flavons and  extensively studied three non-identical $M_S$ structures separately in NH as well as in IH mode.

In conclusion, the low scale MES mechanism is analyzed in this work. This model can also be used to study the connection between effective mass in neutrinoless double beta decay in a wider range of sterile neutrino mass from $eV$ to few $keV$. Study of $keV$ scale sterile neutrino can be a portal to explain origin of dark matter and related cosmological issues in this MES framework.

\section{ACKNOWLEDGEMENTS}
We would like to thank Department of Computer Science and Engineering, Tezpur University for giving us the full access to The High performance computing facility (a Joint project by Tezpur university and C-DAC Pune) for our simulation work. 

\appendix

\section{Product rules and Vacuum Alignment under $A_4$}\label{a4p}

$A_4$, the symmetry group of a tetrahedron, is a discrete non-Abelian group of even permutations of four objects. It has 12 elements with four irreducible representations: three one-dimensional and one three-dimensional which are denoted by $\bf{1}, \bf{1'}, \bf{1''}$ and $\bf{3}$ respectively. Cube root of unity is defined as $\omega=exp(i\frac{2\pi}{3})$, such that $1+\omega+\omega^2=0$. $A_4$ can be generated by two basic permutations $S$ and $T$ given by $S=(4321)$ and $T=(2314)$ (For a generic (1234) permutation). One can check immediately as,
$$S^2=T^3=(ST)^3=1$$.
The irreducible representations for the $S$ and $T$ basis are different from each other. We have considered the $T$ diagonal basis as the charged lepton mass matrix is diagonal in our case. Their product rules are given as,
$$ \bf{1} \otimes \bf{1} = \bf{1}; \bf{1'}\otimes \bf{1'} = \bf{1''}; \bf{1'} \otimes \bf{1''} = \bf{1} ; \bf{1''} \otimes \bf{1''} = \bf{1'}$$
$$ \bf{3} \otimes \bf{3} = \bf{1} \otimes \bf{1'} \otimes \bf{1''} \otimes \bf{3}_a \otimes \bf{3}_s $$
where $a$ and $s$ in the subscript corresponds to anti-symmetric and symmetric parts respectively. Denoting two triplets as $(a_1, b_1, c_1)$ and $(a_2, b_2, c_2)$ respectively, their direct product can be decomposed into the direct sum mentioned above as,
\begin{equation}
\begin{split}\label{a4r}
& \bf{1} \backsim a_1a_2+b_1c_2+c_1b_2\\
& \bf{1'} \backsim c_1c_2+a_1b_2+b_1a_2\\
& \bf{1''} \backsim b_1b_2+c_1a_2+a_1c_2\\
&\bf{3}_s \backsim (2a_1a_2-b_1c_2-c_1b_2, 2c_1c_2-a_1b_2-b_1a_2, 2b_1b_2-a_1c_2-c_1a_2)\\
& \bf{3}_a \backsim (b_1c_2-c_1b_2, a_1b_2-b_1a_2, c_1a_2-a_1c_2)\\
\end{split}
\end{equation}

 Here we will investigate the problem of achieving the VEV alignment of the two flavons. We will take the minimization potential and try to solve them simultaneously. In general, total potential will be consisting of the contribution from the field $\zeta$ and $\varphi$ and their mutual interaction. However, interaction among the fields are forbidden by the discrete charges.
 The total potential will be,
 \begin{equation}
 V=V(\zeta)+V(\varphi)+V_{int.},
  \end{equation}
with,
$$ V(\zeta)=-m_1^2(\zeta^{\dagger}\zeta)+\lambda_{1}(\zeta^{\dagger}\zeta)^2$$ and
$$V(\varphi)=-m_2^2(\varphi^{\dagger}\varphi)+\lambda_{1}(\varphi^{\dagger}\varphi)^2$$
$V_{int.}$ term will not appear in our case.\\
The triplet fermions will have the form,
\begin{equation}
\begin{split}
&\langle \zeta\rangle = (\zeta_{1},\zeta_{2},\zeta_{3}),\\
&\langle \varphi\rangle=(\varphi_{1},\varphi_{2},\varphi_3)
\end{split}
\end{equation}
Using the $A_4$ product rules from equ. \eqref{a4r} , the potential for $\varphi$ will take the form,
\begin{equation}
\begin{split}
V(\varphi)=&-\mu_2^2(\varphi_{1}^{\dagger}\varphi_{1}+\varphi_2^{\dagger}\varphi_3+\varphi_3^{\dagger}\varphi_2)\\
&+\lambda_2[(\varphi_{1}^{\dagger}\varphi_{1}+\varphi_2^{\dagger}\varphi_3+\varphi_3^{\dagger}\varphi_2)^2+(\varphi_{3}^{\dagger}\varphi_{3}+\varphi_2^{\dagger}\varphi_1+\varphi_1^{\dagger}\varphi_2)\times(\varphi_{2}^{\dagger}\varphi_{2}+\varphi_1^{\dagger}\varphi_3+\varphi_3^{\dagger}\varphi_1)\\
&+(2\varphi_{1}^{\dagger}\varphi_{1}-\varphi_2^{\dagger}\varphi_3-\varphi_3^{\dagger}\varphi_2)^2+2(2\varphi_{3}^{\dagger}\varphi_{3}-\varphi_1^{\dagger}\varphi_2-\varphi_2^{\dagger}\varphi_1)\times(2\varphi_{2}^{\dagger}\varphi_{2}-\varphi_3^{\dagger}\varphi_1-\varphi_1^{\dagger}\varphi_3)]\\
\end{split}
\end{equation}
Taking the derivative w.r.t. $\varphi_{1}, \varphi_{2}$ and $\varphi_3$ and equate it to zero gives us the minimization condition for the potential. Three equations are solved simultaneously and various solutions are found out as,
\begin{enumerate}
	\item $\varphi_1\rightarrow\frac{\mu_2}{\sqrt{10\lambda_2}},\varphi_2\rightarrow0,\varphi_3\rightarrow0\Rightarrow\langle\varphi\rangle=\frac{\mu_2}{\sqrt{10\lambda_2}}(1,0,0)$;
	\item $\varphi_1\rightarrow\frac{\mu_2}{2\sqrt{3\lambda_2}},\varphi_2\rightarrow\frac{\mu_2}{2\sqrt{3\lambda_2}},\varphi_3\rightarrow\frac{\mu_2}{2\sqrt{3\lambda_2}}\Rightarrow\langle\varphi\rangle=\frac{\mu_2}{2\sqrt{3\lambda_2}}(1,1,1)$;
	\item $\varphi_1\rightarrow\frac{2\mu_2}{\sqrt{51\lambda_2}},\varphi_2\rightarrow-\frac{\mu_2}{\sqrt{51\lambda_2}},\varphi_2\rightarrow-\frac{\mu_2}{\sqrt{51\lambda_2}}\Rightarrow\langle\varphi\rangle=\frac{\mu_2}{\sqrt{51\lambda_2}}(2,-1,-1)$.
\end{enumerate}
Similar solutions will be generated for the $\zeta$ field also. We have used the first set of solution in order to generate charged lepton mass matrix and solution (2) and (3) for the generation of the Dirac neutrino mass matrix within NH and IH mode respectively.
 \section{Comparison among extended seesaw}\label{apa}
\begin{itemize}
	\item 
	{\bf Inverse Seesaw (IS) :} The neutral mass matrix in the basis ($\nu_{L}, \nu_{R}^c,S^c$) takes the form
	\begin{equation}
	M_{\nu}^{IS}=
	\begin{pmatrix}
	0&M_D&0\\M_{D}^{T}&0&M_{S}^{T}\\0&M_{S}&\mu 
	\end{pmatrix}.
	\end{equation}
	In this case, $\mu,M_D<<M_S$ and this scenario has been termed as the inverse seesaw. Typically here $n(\nu_L)=n(\nu_R)=S=3$ however $n(\nu_R)=n(S)=2$ is also viable.
	\item {\bf Extended Inverse Seesaw(EIS) :} Here the mass matrix is extended within the same basis as ISS with an extra mass term 
	\begin{equation}
	M_{\nu}^{EIS}=
	\begin{pmatrix}
	0&M_D&0\\M_{D}^{T}&\mu_1&M_{S}^{T}\\0&M_{S}&\mu_2 
	\end{pmatrix}.
	\end{equation}
	This is an extended inverse seesaw mechanism for $\mu_i,M_D<<M$. This scenario is more or less similar to the inverse seesaw mechanism and works with same number of states.
	\item {\bf Extended Seesaw (ES) :} The mass matrix of extension of type-I seesaw is already defined in equation \ref{7b7} as,
	\begin{equation}
	M_{\nu}^{ES}=
	\begin{pmatrix}
	0&M_D&0\\M_{D}^{T}&\mu&M_{S}^{T}\\0&M_{S}&0 
	\end{pmatrix}.
	\end{equation}
\end{itemize}

In spite of the fact that Extended Seesaw looks like the Extended Inverse seesaw, however this is not an inverse seesaw mechanism. For this extended seesaw scenario as $\mu\gg M_{S}>M_{D}$, analogous to the type-I seesaw. We have considered the number of states as $n(\nu_{L})=n(\nu_{R})=3$ and $n(S)=1$ and this is the minimal extended seesaw picture. Even though three different $S$ were considered within our study, but they independently generate three different $M_S$ matrices. Hence, our model still behaves as a minimal extended seesaw. The RH neutrino mass scale is near to the GUT scale for MES(or ES), whereas for the inverse seesaw case, the RH mass scale is much smaller as compared to the  earlier one, hence the lepton number violating scale also.

\bibliographystyle{apsrev}

\end{document}